\documentstyle[tighten,preprint,amsfonts,eqsecnum,aps,prd]{revtex}
\begin{document}
\draft

\hyphenation{
mani-fold
mani-folds
geo-metry
geo-met-ric
}


\def\bar{\overline}             
\def\hat{\widehat}              

\def\tr{{\rm tr}}                    
\def\BbbR{{\Bbb R}}
\def\BbbD{{\Bbb D}}

\def\half{{1\over2}}
\def\casehalf{{\case{1}{2}}}

\def\tstar{t_*}

\def\sothree{\hbox{SO(3)}}
\def\sofour{\hbox{SO(4)}}
\def\ofour{\hbox{O(4)}}
\def\sutwo{\hbox{SU(2)}}
\def\utwo{\hbox{U(2)}}
\def\uone{\hbox{U(1)}}

\def\bfx{{\bf x}}

\def\hilbert{{\sf H}}

\def\Psinb{\Psi_{\rm NB}}

\def\desitter{de~Sitter}
\def\tnds{Taub-NUT-de~Sitter}

\def\Mtilde{{\widetilde M}}
\def\Atilde{{\tilde A}}

\def\ballthreec{{\overline B}^3}
\def\ballfourc{{\overline B}^4}

\def\RPthree{{\Bbb RP}^3}
\def\RPfour{{\Bbb RP}^4}
\def\Reals{{\Bbb R}}

\def\crosscapfourc{{\overline M}_\otimes}



\preprint{\vbox{\baselineskip=12pt
\rightline{PPG98--123}
\rightline{SU--GP--98/5--3}
\rightline{gr-qc/9805101}}}

\title{Instantons and unitarity in quantum cosmology
\\
with fixed four-volume}
\author{Alan Daughton\footnote{Electronic address: 
daughton@nuclecu.unam.mx}}
\address{
Department of Physics,
Syracuse University,
Syracuse, 
New York 13244--1130, 
USA
\\
and
\\
Instituto de Ciencias Nucleares, 
UNAM,
A.~Postal 
70-543, 
D.F. 04510, 
Mexico\footnote{Present address.} 
}
\author{Jorma Louko\footnote{
Electronic address:
louko@aei-potsdam.mpg.de
}}
\address{
Department of Physics,
Syracuse University,
Syracuse, 
New York 13244--1130, 
USA
\\
and
\\
Department of Physics,
University of
Wisconsin--Milwaukee,
\\
P.O.\ Box 413,
Milwaukee, Wisconsin 53201, 
USA
\\
and
\\
Department of Physics,
University of Maryland,
College Park,
Maryland 20742--4111,
USA
\\
and
\\
Max-Planck-Institut f\"ur Gravitations\-physik,
Schlaatzweg~1,
D--14473 Potsdam,
Germany\footnote{Present address.} 
}
\author{Rafael D. Sorkin\footnote{Electronic address:
sorkin@suhep.phy.syr.edu}}
\address{
Department of Physics, 
Syracuse University,
Syracuse, 
New York 13244--1130, 
USA\footnote{Permanent address.}
\\
and
\\
Instituto de Ciencias Nucleares, 
UNAM,
A.~Postal 70-543, 
D.F. 04510, 
Mexico
}
\date{Revised version, August 1998. 
Published in {\it Phys.\ Rev.\ \rm D \bf 58} (1998) 084008.}
\maketitle

\begin{abstract}%
We find a number of complex solutions of the source-free Einstein
equations in the so-called unimodular version of general relativity,
and we interpret them as saddle points yielding estimates of a
gravitational path integral taken over a space of almost everywhere
Lorentzian metrics on a spacetime manifold with topology of the
``no-boundary'' type.  Within this interpretation, we address the
compatibility of the no-boundary initial condition with the
definability of the {\em quantum measure\/}, which reduces in this
setting to the normalizability and unitary evolution of the
no-boundary wave function $\psi$.  We consider three spacetime
topologies, $\Reals^4$, $\RPfour\#\Reals^4$, and
$\Reals^2{\times}T^2$.  (The corresponding truncated
manifolds-with-boundary are respectively the closed 4-dimensional
disk or ball, the closed 4-dimensional cross-cap, and the product of
the two-torus with the closed two-dimensional disk.)  The first two
topologies we investigate within a Taub minisuperspace model with
spatial topology~$S^3$, and the third within a Bianchi type I
minisuperspace model with spatial topology~$T^3$.  In each of the
three cases there exists exactly one complex solution of the
classical Einstein equations (or combination of solutions) that, to
the accuracy of our saddle point estimate, yields a wave function
compatible with normalizability and unitary evolution.  The
existence of such solutions tends to bear out the suggestion that
the unimodular theory is less divergent than traditional Einstein
gravity.  In the Bianchi type I case, moreover, the distinguished
complex solution is approximately real and Lorentzian at late times,
and appears to describe an explosive expansion from zero size at
$T=0$.  In this connection, we speculate that a fully normalizable
$\psi$ can result only from the imposition of an explicit short
distance cutoff.  (In the Taub cases, in contrast, the only complex
solution with nearly Lorentzian late-time behavior yields a wave
function that is normalizable but evolves nonunitarily, with the
total probability increasing exponentially in the unimodular
``time'' in a manner that suggests a continuous creation of new
universes at zero volume.)  The issue of the stability of these results
upon the inclusion of more degrees of freedom is raised.
\end{abstract}


\narrowtext

\section{Introduction}
\label{sec:intro}

In formulating the gravitational functional integral
on a compact manifold with boundary,
\begin{equation}
\int {\cal D} g\, \exp \left[iS(g)\right]
\ \ ,
\end{equation}
one may choose to limit the geometries $g$ that enter the sum by specifying
a fixed value for the total 4-volume 
\cite{sorkin-talk,goa,forks}.  
If $S(g)$ is the Einstein-Hilbert
action, this restriction produces a theory whose classical limit is
equivalent to Einstein's theory with a cosmological constant, the only
difference being that the cosmological constant arises as a constant
of integration and not as a prescribed parameter in the action
\cite{sorkin-talk,goa,forks,einstein,vanderBij,%
wilczek,zee,buchdragon,%
weinberg,unruh,unruh-wald,henn-teitel,brown-york,%
bombelli-banff,kuchar,casta-lom}.  
This theory is often called unimodular gravity, owing to the fact that one
can alternatively derive it by imposing the coordinate condition
$\sqrt{-g}=1$ in the action prior to variation.

In our view, the motivation for a unimodular modification of gravity
is threefold \cite{sorkin-talk,forks}.  First, it may help explain why
the cosmological constant can be so small.  Second, it is suggested by
analogy with the structure of nonrelativistic quantum mechanics.
Third, based on this analogy, it can be expected to improve the
convergence of certain expressions that arise in the computation of
the quantum measure of a set of histories ({\em i.e.}, of
4-geometries).  This third motivation is the most relevant to the
present paper.

In a histories framework for quantum theory
(see Refs.\ 
\cite{sorkin-talk,forks,feyn,griffiths,%
sorkin-einstein,ishamhist,qmqmt,hartlehist,drexel}
and references therein), the quantum measure $\mu$ plays a role
analogous to that 
played by the classical probability-measure in a classically stochastic
process such as diffusion.  Within a histories framework, no wave function
ever need be introduced, but it is often convenient to do so, because 
$\mu$ can often be computed as $||\psi||^2$ for a suitable $\psi$. In
unimodular quantum gravity in particular, one can introduce a wave
function on 3-geometries by summing over 4-geometries with fixed
4-volume \cite{sorkin-talk}, and the resulting $\psi$ will depend on the
4-volume $T$ of the spacetimes that enter into the sum.  Now, if the
relation between $\psi$ and the quantum measure $\mu$ in quantum gravity
is like that in ordinary quantum mechanics, then, in order that $\mu$ be
well-defined, it is necessary first of all that $||\psi||^2$ be finite,
and secondly that it be independent of 4-volume, 
{\em i.e.}, 
that the ``evolution'' of $\psi$ with ``time'' $T=V$ be unitary. 
These are the principal questions that we explore in the present paper.

In the present paper, our considerations will be based on a Lagrangian
formulation, both for its relative simplicity and because it is the most
suitable formulation for dealing with the type of topology change that a
``big bang'' cosmology entails.  Nevertheless, it may be of some interest
to sketch here how the unimodular assumption manifests itself in
Hamiltonian versions of gravity.  To understand what happens to the
constraints, it is useful to think in terms of the path integral: the
condition of fixed spacetime volume removes one degree of freedom from
the permissible deformations of the final hypersurface, and this in turn
eliminates one of the infinity of Hamiltonian constraints that are
present in the conventional formulation, or rather converts it into a
Schr{\"o}dinger equation expressing the dependence of $\psi$ on~$T$.  In
one particular Hamiltonian scheme for unimodular gravity with closed
spatial hypersurfaces, this works out in more detail as follows.  The
theory contains a pair of canonically conjugate fields that are not
present in the canonical formulation of conventional Einstein
gravity. One of the new fields specifies the value of the cosmological
constant, while the conjugate field carries the information about the
spacetime volume bounded by the initial and final spacelike
hypersurfaces.  Dirac quantization of this Hamiltonian theory leads, in
addition to a set of constraint equations, to a Schr\"odinger-type
equation in which the ``time'' variable can be identified as the
four-dimensional spacetime volume.  One therefore would expect to adopt
a Schr\"odinger-type Hilbert space in which the Hamiltonian would be a
selfadjoint
operator, and the wave function would evolve unitarily in
``unimodular time.''  This ``unfreezing'' of the wave function raised
hopes that the interpretational issues of quantum gravity, especially
regarding time\cite{ash-time,kuchar-winn,isham-time}, might be more
easily tractable within the unimodular theory than in the conventional
theory.  However, from a histories standpoint, no ``problem of time'' is
evident, and the role of unimodularity would seem to be more technical
than interpretational in nature.  For some further discussion, see
Refs.\ \cite{kuchar,kuchar-winn}.

In this paper we explore the implications of unimodular gravity for
quantum cosmology.  Specifically we explore its implications for
no-boundary initial conditions of the sort proposed by 
Hartle and Hawking 
\cite{hawking-vatican,hartle-hawking,hawking-npb}, 
Linde 
\cite{linde-zetf,linde-ncim,linde-repprog,linde-mezh,linde2-mezh}, 
and Vilenkin 
\cite{vile1,vile2a,vile2b,vile3,vile-erice,vile-discord}, 
and more generally,
for the framework for topology change 
set out in Refs.\ \cite{forks,victoria,rafael-arvind}.

The condition we impose 
is that spacetime be a 4-manifold $M$ that
is compact toward the past, with empty initial boundary.\footnote
{This condition can be made precise in the language of Morse theory: 
$M$~should admit a ``height function'' $h\ge{0}$ 
with the property that
$h^{-1}([0,r])$ is compact 
and $h^{-1}([0,r))$ is boundary-free
for all real $r>0$.}
When truncated toward the future in order to compute the quantum measure
\cite{sorkin-talk,sorkin-einstein}, $M$ will therefore acquire a future
boundary $M_1$ that is 
{\em closed\/} 
in the technical sense of being compact
without boundary.\footnote%
{The manifolds $M$ we consider in the present paper are all such that $M_1$
may be assumed to be a smooth 3-manifold.  In cases where topological
transitions are not limited to the ``moment of birth'' of the universe,
the level surfaces of a Morse height function are not all manifolds.  This
suggests that, in computing the quantum measure for such spacetimes, one
might want to consider wave functions defined on some sort of
correspondingly generalized 3-geometries.}
A wave function obtained from a unimodular path integral over such a
truncated manifold will have as arguments the 4-volume $T$ and the
induced 3-geometry on the 3-manifold $M_1$.

Precisely what kinds of
metrics are to be integrated over --- 
or indeed, whether it is possible to define consistently a gravitational
functional integral at all in a continuum theory --- 
is a poorly understood issue
\cite{rafael-arvind,forks,hawkingCC,hall-hartle-con,jjhlou3} 
to which we shall return in section~\ref{sec:discussion}.  
For now, we mention only that the point of view we adopt in the following
is that the path integral is originally over (almost everywhere)
Lorentzian metrics, and any complex metrics one considers have
meaning only insofar as they yield approximations to such a Lorentzian
path integral.  
In the main part of the paper, we will simply assume that the path
integral can analyzed in the crudest possible saddle-point
approximation; and we will find the complex classical solutions for the
unimodular theory, without 
attempting to control even 
the semiclassical prefactors.
As the unimodular boundary condition requires the Lorentzian 4-volume to
be real, 
the saddle-point geometries are in general 
{\em necessarily complex\/},
although saddle-points with Lorentzian signature will 
be seen to exist in certain special cases.  We emphasize that this
condition of real 4-volume excludes from our framework any geometry with
purely Euclidean signature.

As explained above, the crucial consistency conditions for the quantum
measure to be defined (and therefore for the path integral to lead to
meaningful predictions) are that the path integral give a wave
function that, with respect to a suitable measure, is square
integrable and evolves unitarily.  We investigate these features
within two spatially homogeneous minisuperspace models: the Taub model
(Bianchi type IX plus an additional $\uone$ symmetry) with $S^3$
spatial topology, and Bianchi type~I with $T^3$ spatial topology and a
certain additional discrete symmetry.  As the (truncated)
no-boundary 4-manifolds,
we consider the closed 4-ball $\ballfourc$ and the closed
4-dimensional cross-cap $\RPfour \# \ballfourc$ in the Taub model, and
the closed disk times the two-torus in Bianchi type~I\null. In all
cases, finding the no-boundary saddle points reduces to solving a
simple algebraic equation.  Having found the saddle points, 
we first ask whether a
saddle-point estimate to the path integral is compatible with a
normalizable wave function, for any choice of saddle point(s).  If yes,
we then ask whether the corresponding wave function evolves unitarily,
to the approximation in question.  Also, we ask whether any of the
saddle-point geometries are approximately Lorentzian at late times.  
Finally, we ask how the saddle
point wave function behaves at $T=0$, and in particular, whether this
behavior seems compatible with the picture of a universe expanding
from zero size that is implicit in the no-boundary topology.

In the Taub model, for both of our 4-manifolds, we find a unique saddle
point that, to the accuracy of our estimate, is compatible
with both normalizability and unitary evolution.  This bears out the
suggestion \cite{sorkin-talk,goa,forks}
that the unimodular theory is less divergent than traditional
Einstein gravity, and tells us in each case
what the approximate behavior of the
wave function  must be if the quantum measure is indeed well defined.
Interestingly, this saddle point remains always in the quantum era,
never making a spontaneous transition to classical behavior.
In addition there is (for both 4-manifolds) a unique saddle point that
is compatible with normalizability and 
does make a transition to classical behavior.
However, the wave function corresponding to
this saddle point turns out {\em not\/} to evolve unitarily: instead,
probability is being injected into the configuration space at a rate that
is exponentially increasing in the unimodular time. This injection appears
to take place at a boundary of the
configuration space, in a manner reminiscent of the tunneling boundary
conditions advocated by 
Linde 
\cite{linde-zetf,linde-ncim,linde-repprog,linde-mezh,linde2-mezh} 
and Vilenkin 
\cite{vile1,vile2a,vile2b,vile3,vile-erice,vile-discord}. 
Physically, such
an injection can perhaps be interpreted as a continuous creation of new
``branch universes,'' all stemming from a single root.  

In the Bianchi type I model, the unique saddle point that is compatible
with normalizability turns out to be compatible also with unitary
evolution. Further, the saddle-point geometries are, at late times,
nearly Lorentzian, isotropically expanding universes.  Thus, this saddle
point exhibits many features normally regarded as desirable for quantum
cosmology.

The plan of the paper is as follows.  In section
\ref{sec:taub-in-unimod} we introduce the unimodular Taub
minisuperspace model and the unimodular positive curvature Friedmann
model, which arises as the isotropic specialization of the Taub model.
Sections \ref{sec:taub-b4} and \ref{sec:taub-cross} discuss the Taub
no-boundary saddle points when the 4-manifold is respectively the
closed 4-ball and the closed cross-cap, and section
\ref{sec:fried-restr} discusses the truncation of these no-boundary
analyses to the Friedmann model. The Bianchi type I model is analyzed
in section~\ref{sec:bianchiI}.
Our results are summarized and discussed in
section~\ref{sec:discussion}. 

We use throughout units such that $c=\hbar=1$, but we keep Newton's
constant~$G$. A~metric with signature
$({-}{+}{+}{+})$ is called Lorentzian, and a metric with signature
$({+}{+}{+}{+})$ Riemannian.

\section{Taub minisuperspace in the unimodular theory}
\label{sec:taub-in-unimod}

In this section we describe the Taub minisuperspace model in the
unimodular theory. Subsection \ref{subsec:taub-in-unimod-gen} presents the
general case, with two independent scale factors. The truncation to the
positive curvature Friedmann model is outlined in
subsection~\ref{subsec:friedmann-in-unimod}.
We assume throughout this section that the spacetime is everywhere
Lorentzian and has topology $\BbbR \times S^3$.
The boundary conditions needed to express the no-boundary topologies we
consider will be introduced in 
section~\ref{sec:taub-b4}.

\subsection{The general Taub model}
\label{subsec:taub-in-unimod-gen}

The Taub family of metrics can be written as 
\cite{ryan-shepley,jantzen-cmp}
\begin{equation}
   ds^2 = 
    \sigma^2 \left\{- N^2 dt^2 + \case{1}{4} a^2 {(\omega^1)}^2
    + \case{1}{4} b^2 \! \left[ {(\omega^2)}^2 + {(\omega^3)}^2 \right]
   \right\}
  \ \ ,
  \label{taubmetric}
\end{equation}
where $a$, $b$, and $N$ are functions of~$t$, and the $\omega^i$ are the
usual left-invariant one-forms on $\sutwo$, satisfying
\begin{equation}
  d\omega^i = -\half \epsilon^i{}_{jk} \omega^j \wedge \omega^k
  \ \ .
\end{equation}
We use conventions in which the exterior derivative and the wedge
product 
are
$(d\omega)_{ab}=\partial_a\omega_b-\partial_b\omega_a$
and $(\omega\wedge\phi)_{ab}= \casehalf
\left(\omega_a\phi_b-\omega_b\phi_a\right)$, and we have extracted in
(\ref{taubmetric}) the overall factor $\sigma^2 := 2G/3\pi$ for
numerical
convenience.  As $\sigma$ has the dimension
of length, we can take $a$, $b$, $N$, $t$, and $\omega^i$ to be
dimensionless. In the special case $a=b$, the spatial sections are
round 3-spheres with radius of curvature $\sigma a$.

The spacetime topology is
$\BbbR \times \sutwo
\simeq 
\BbbR \times S^3$, 
and the spacetime isometry group is 
that of the constant $t$ hypersurfaces, 
$\utwo \simeq \sutwo_L \times \uone_R / {\Bbb Z}_2$.  The $\sutwo$ factor
comes from the invariance of (\ref{taubmetric}) under the left action of
$\sutwo$ on itself, and the further $\uone$ isometry (acting on the right)
expresses the equality of
the coefficients of ${(\omega^2)}^2$ and~${(\omega^3)}^2$.

Inserting the metric (\ref{taubmetric}) into the gravitational
action-integral, 
\begin{equation}
  S = {1\over 8\pi G} \left[ 
  \int (\case{1}{2}R - \Lambda) dV 
 \ + \ 
 \oint \tr K
 \right]
\ \ ,
\label{S-grav}
\end{equation}
with (bare) cosmological constant~$\Lambda$, yields the
minisuperspace action-integral
\begin{equation}
  S = \case{1}{6}
      \int d\tau
      \left[
      - a \left({db\over d\tau}\right)^2 
      - 2b {da\over d\tau} {db\over d\tau} 
      + 4a - a^3 b^{-2} 
      - 3\lambda a b^2 
      \right]
\ \ ,
\label{taub-conv-action}
\end{equation}
where we have introduced the dimensionless proper time parameter $\tau$ by
$d\tau:=Ndt$ and written $\lambda := \case{1}{3} \sigma^2 \Lambda$.  The
true proper time is~$\sigma\tau$.

Given the action-integral, it is 
easy 
to derive the classical
equations of motion in both the unimodular and non-unimodular theories.
In the non-unimodular theory, the classical equations of motion result from
making arbitrary variations of $S$ that fix the metric on the boundary.
Because the ansatz (\ref{taubmetric}) expresses invariance under a compact
symmetry group, it suffices to consider variations of the 
parameters $a$, $b$, and~$\tau$.  
The condition of fixed boundary metric is then equivalent
to fixing $a$ and $b$ (but not $\tau$) at the endpoints.\footnote%
{In an action that retains $t$ and~$N$, the values of $t$ at the
  endpoints can be fixed, and the equation that results from variation
  with respect to $N$ is equivalent to the equation obtained by
  varying (\ref{taub-conv-action}) with respect to $\tau$ at the
  endpoints.}

The general solution to the variational equations can be written in
the gauge $Na=1$ as
\begin{mathletters}
\label{taub-class-sol}
\begin{eqnarray}
b^2 &=& A^2 + t^2/A^2
\ \ ,
\label{taub-class-sol-b}
\\
a^2b^2 &=& A^2 \left\{ 4 \left( A^2 - t^2/A^2 \right)
+ 3 \lambda \left[ t^4/(3A^4) + 2t^2 - A^4 \right] \right\}
+ ABt
\ \ ,
\label{taub-class-sol-aabb}
\\
N &=& 1/a
\ \ ,
\end{eqnarray}
\end{mathletters}%
where $A>0$ and $B\in\BbbR$ are the two constants of integration.
This metric covers the spatially homogeneous region of the \tnds\ 
solution \cite{exact}.  In the notation of Ref.\ \cite{exact}, the
Taub-NUT ``charge'' $l$ and ``mass-parameter'' $m$ are given by
$l=\half\sigma A$ and $m={1\over4}\sigma B$. 

For the purposes of the unimodular theory, it is convenient to write
the action-integral in a slightly different form.  First of all, we
may set $\Lambda$ to zero in (\ref{S-grav}) without loss of
generality, because it influences neither the classical solutions nor
the quantum measure. (We obtain the classical
equations of motion
by varying $S$ subject to the condition that the total 4-volume is
fixed.)  Second, because of the special role played by the spacetime
4-volume, it is useful to adopt it as our time coordinate (see for
example Ref.\ \cite{sorkin-talk}).  For numerical convenience, we may
use the dimensionless 4-volume parameter $T$ 
defined by $dT=ab^2d\tau$: the
4-volume bounded by the hypersurfaces $T=T_1$ and $T=T_2$, with
$T_1<T_2$, is then $2\pi^2 \sigma^4 (T_2-T_1)$.  Writing
(\ref{taub-conv-action}) in terms of~$T$ (with $\lambda$ set to zero)
we obtain the action-integral in the form
\begin{equation}
 S =  \case{1}{6} \int dT 
  \left( - a^2 b^2 b'^2 - 2a b^3 a' b' 
  + 4 b^{-2} - a^2 b^{-4} \right)
\ \ , 
\label{taub-uaction}
\end{equation}
where the prime denotes derivative with respect to~$T$. In the
variation of~(\ref{taub-uaction}), fixing the total spacetime
volume is equivalent to fixing the difference between the
initial and final values of~$T$, and as the integrand does not involve
$T$ explicitly, this is further equivalent to fixing individually the
initial and final values of~$T$. The general solution to the resulting
variational equations is precisely~(\ref{taub-class-sol}), but
$\lambda$ emerges now as an integration constant proportional to the
``unimodular energy,'' and the general solution contains thus three
constants of integration. 

It is convenient to replace $a$
and $b$ by the coordinates (cf.\ \cite{sorkin-talk,rayrev}) 
\begin{equation}
\begin{array}{rcl}
u &:=& a^2 b
\ \ ,
\\
v &:=& b^3
\ \ , 
\end{array}
\label{uv-def}
\end{equation}
which represent spatial volumes rather than lengths.  In these coordinates,
the action-integral (\ref{taub-uaction}) simplifies to 
\begin{equation}
 S = \int dT \left(
   - \case{1}{18} u'v' 
   - \case{1}{6}u v^{-5/3} 
   + \case{2}{3}v^{-2/3}
        \right)
\ \ .
\label{S-taub-uv}
\end{equation}
The configuration space of the theory is therefore the future quadrant
of a (1+1)-dimensional Minkowski space, and $(u,v)$ forms a pair of
positive-valued null coordinates.  
(Notice that in this system, the ``radial coordinate'' $\sqrt{uv}$
represents the spatial volume.) 
The Hamiltonian operator
corresponding to (\ref{S-taub-uv}) is
\begin{equation}
{\hat H} :=
  18 {\partial^2 \over \partial u \partial v}
  + \case{1}{6} u v^{-5/3} - \case{2}{3} v^{-2/3}
\ \ , 
\label{taub-Hamiltonian}
\end{equation}
where we have chosen to factor-order the kinetic part of the
Hamiltonian as the D'Alem\-bertian (or ``Laplace-Beltrami operator'')
of the metric associated to the kinetic terms in the Lagrangian with
$T$ used as time-parameter.  Notice that a change of time-parameter
would change this metric and consequently change what we would call
the D'Alembertian.  Conversely, the combination of 4-volume time with
Laplace-Beltrami ordering with respect to that time provides a
universal factor-ordering for all spatially homogeneous cosmological
models\cite{mike-rds}.  The Hilbert space inner product that goes
naturally with a metric is the $L^2$ integral with respect to the
metric's volume element, and the Laplace-Beltrami 
operator is formally selfadjoint with respect to this
inner product, as is easily seen in general.  In the present case this
inner product is
\begin{equation}
  (\psi_1,\psi_2) :=
  \int_{u>0 \atop v>0}
  du \, dv \>
  \overline{\psi_1} \psi_2
\ \ .
\label{ip}
\end{equation}
As ${\hat H}$ is real, it has selfadjoint extensions
by von Neumann's theorem (Ref.\ \cite{reed-simonII}, Theorem~X.3).
Choosing one such extension specifies a unitary evolution in the
Hilbert space of square integrable functions~$\psi$.

\subsection{Friedmann truncation}
\label{subsec:friedmann-in-unimod}

The Taub model can be truncated to its isotropic special
case by setting $a=b$. The metric (\ref{taubmetric}) becomes then the
metric of the positive curvature Friedmann model,
\begin{equation}
ds^2 = \sigma^2 \left[
- N^2 dt^2 + a^2 d\Omega_3^2 \right]
\ \ ,
\label{friedmetric}
\end{equation}
where $d\Omega_3^2$ is the metric of the unit
3-sphere.\footnote{Our conventions in the Taub model were chosen
  so as to make the Friedmann metric (\ref{friedmetric}) agree with
  the conventions of Refs.\ \cite{hartle-hawking,hawking-npb}.} 
The isometry group of the constant $t$ hypersurfaces is~$\ofour$. 
The  action (\ref{taub-uaction}) becomes
\begin{equation}
  S = \casehalf \int dT 
  \left( -a^4 a'^2 + a^{-2} \right)
  \ \ . 
\label{S-friedmann-unimod}
\end{equation}
The Hamiltonian operator corresponding to (\ref{S-friedmann-unimod})
is 
\begin{equation}
{\hat H}
  := \half 
     \left[ 
       9 \left( {d^2 \over dx^2} \right) - x^{-2/3} 
     \right]
  \ \ ,
\end{equation}
where $x:= a^3$ 
and we have again chosen the Laplace-Beltrami factor ordering belonging
to $T$ as time parameter. 
The matching inner product is
\begin{equation}
  (\psi_1,\psi_2) := \int_0^\infty dx \, \overline{\psi_1} \psi_2
  \ \ .
\label{fip}
\end{equation}
and ${\hat H}$~is formally 
selfadjoint 
with respect to it. The 
selfadjoint 
extensions of ${\hat H}$ are specified 
by the boundary condition
$\cos(\theta) \psi - \sin(\theta) d\psi/dx =0$ at $x=0$, 
where the parameter $\theta$ 
satisfies $0\le \theta < \pi$ 
(Ref.\ \cite{reed-simonII}, Theorems X.8 and X.10). 
Choosing the value of $\theta$ specifies a unitary quantum evolution.

\section{Taub no-boundary saddle points for $\ballfourc$ topology}
\label{sec:taub-b4}

In this section we discuss the Taub no-boundary 
saddle points and wave functions when the
4-manifold is the closed 4-dimensional 
ball~$\ballfourc$ (often called disk ${\Bbb D}^4$).
In subsection \ref{subsec:taub-b4-geoms} 
we find the saddle-point 
geometries, 
and in subsection \ref{subsec:taub-b4-wavefuncs} 
we discuss the saddle-point estimates
to the wave function.

As discussed in the Introduction, we introduce a wave function via the
(formal) no-boundary path integral
\begin{equation}
  \Psinb(a,b;T) = \int {\cal D} g \, \exp \left(iS\right)
\ \ .
\label{taub-nb-integral}
\end{equation}
where the domain of integration is some appropriate class of 
Lorentzian, or almost Lorentzian, 4-geometries, possibly with a
short-distance cutoff.
In the histories entering the integral, the ``final'' values of
the scale factors and the total elapsed volume parameter $T$ 
are required to coincide with the arguments of the wave function~$\Psinb$;
in particular, $T$ in $\Psinb$ is assumed always positive. 

In this paper we only consider $\Psinb$ in the saddle-point estimate:
\begin{equation}
  \Psinb \approx \sum_k P_k \exp \left(iS^k\right)
  \ \ ,
\label{taub-nb-estimate}
\end{equation}
where the $S^k(a,b;T)$ are the actions of some subset of the complex
classical solutions on the no-boundary 4-manifold in question.  The
prefactors $P_k$ are assumed to be slowly varying compared with the
exponential factors but otherwise are left unspecified.  Which (if
any) of the saddle point metrics actually contribute to
(\ref{taub-nb-estimate}), given a proper definition of the integral
in~(\ref{taub-nb-integral}), will be left a subject for future work
(see the comments on this issue in 
Ref.\ \cite{forks}).

\subsection{Saddle-point geometries} 
\label{subsec:taub-b4-geoms}

To find the actions $S^k(a,b;T)$  that enter the saddle-point
estimate~(\ref{taub-nb-estimate}), we need on $\ballfourc$ the Taub
solutions with prescribed values of the boundary scale factors and
total 4-volume. The boundary data for the solutions is
Lorentzian, but the saddle-point geometries are allowed to be
complex. 

As a first step, we need the general complex Taub solution to the
unimodular field equations. It is clear that one class of complex Taub
solutions is obtained from the Lorentzian solution
(\ref{taub-class-sol}) by extending the parameters $A$, $B$, and
$\lambda$ to complex values, with $A\ne0$, and making $t$ a
complex-valued function $t(s)$ of a new, real-valued time
coordinate~$s$. We may assume $dt/ds\ne0$.  It can be shown that the
only complex Taub solutions not obtained in this way are obtained in
the same way from the metrics given by
\begin{mathletters}
\label{taub-class-sol-ex}
\begin{eqnarray}
b^2 &=& \pm 2it
\ \ ,
\\
a^2 &=& 
2\lambda t^2 \pm 2it \pm  i D t^{-1} 
\ \ ,
\\
N &=& 1/a
\ \ ,
\end{eqnarray}
\end{mathletters}%
where $D$ and $\lambda$ are complex parameters. The family
(\ref{taub-class-sol-ex}) can be formally recovered from
(\ref{taub-class-sol}) by setting 
$B = 
\pm 8 i A 
\left( 1 - \lambda A^2 \right)
\pm 2iD A^{-3}$, adding $\pm iA^2$ to~$t$, and taking the limit
$A\to\infty$. 

Now, the condition that the solution be defined on $\ballfourc$ means
that $s$ must be interpretable as the radial coordinate of
hyperspherical coordinates on~$\ballfourc$. Without loss of generality
we can take~$s\in[0,1]$, such that $s=0$ occurs at the coordinate
singularity at the origin of the hyperspherical coordinates, and $s=1$
occurs on the boundary. It is then necessary that both $a$
and $b$ vanish at $s=0$.

Let us first concentrate on the family~(\ref{taub-class-sol}).
Writing $t(0)=:t_0$, 
the condition that $a$ and $b$ vanish at $s=0$ implies
\begin{mathletters}
\label{taub-regularity-tnought}
\begin{eqnarray}
t_0 
&=&
i \eta A^2
\ \ , 
\label{taub-tnought}
\\
B
&=&
8 i \eta A 
\left( 1 - \lambda A^2 \right)
\ \ ,
\label{taub-regularity}
\end{eqnarray}
\end{mathletters}%
where $\eta$ is a parameter that may take the values~$\pm1$.  When
equations (\ref{taub-regularity-tnought}) hold, it is straightforward
to verify that the metric is indeed extendible to $s=0$ so that it
defines a solution on~$\ballfourc$.  Thus the smoothness required of
the no-boundary metric is automatic in this case, and does not further
restrict the solution.

To find the action, we recall that when deriving (\ref{taub-uaction})
from (\ref{S-grav}) (with $\Lambda=0$), we assumed the spacetime
topology $\BbbR\times S^3$, for which the boundary term in
(\ref{S-grav}) consists of components both at the initial and final
boundaries.  In the action for the topology~$\ballfourc$,
(\ref{S-grav})~contains a boundary term only at a ``final'' boundary,
and we therefore need to add to (\ref{taub-uaction}) at $s=0$ the
appropriate boundary term, $-\case{1}{6}N^{-1} d(ab^2)/dt$.  This
boundary term however vanishes by virtue
of~(\ref{taub-regularity-tnought}).  
(As earlier with smoothness of
the metric, assuming a no-boundary topology thus turns out to be
indistinguishable, in this case, from simply assuming that the
universe expands from zero diameter.) 
Evaluating (\ref{taub-uaction}) yields then
\begin{equation}
S = i\eta \left[
{\lambda A^4 \rho_*^2 (\rho_*+1)(\rho_*+4) 
\over 6 
(\rho_*+2) }
+ {2A^2 \rho_*(2\rho_*+3) \over 3 (\rho_*+2)}
\right]
\ \ , 
\label{taub-uaction-inter}
\end{equation}
where we have parametrized the value of $t$ at $s=1$ as
$t_0(1+\rho_*)$.

In terms of $A$ and~$\rho_*$, the boundary scale factors and the total
elapsed $T$ are given by 
\begin{mathletters}
\label{taub-alg-system}
\begin{eqnarray}
a^2 &=& 
- { A^2 \rho_* 
\left[ 4 + \lambda A^2 \rho_* (\rho_* + 4) \right]
\over \rho_* + 2}
\ \ ,
\\
b^2 &=& - A^2 \rho_* (\rho_* + 2) 
\ \ ,
\\
T &=& - i \eta A^4 \rho_*^2 
\left( \case{1}{3}\rho_* + 1 \right) 
\ \ .
\end{eqnarray}
\end{mathletters}%
Solving the algebraic equations (\ref{taub-alg-system}) 
for~$A^2$, $\rho_*$, and $\lambda$ yields 
\begin{mathletters}
\label{taub-alg-sol}
\begin{eqnarray}
A^2 &=&
{ 
b^2 {(\epsilon V - 1)}^2 
\over
\epsilon V - 2 }
\ \ ,
\label{taub-alg-sol-A2}
\\
\rho_* &=&
-2 + 
{2 \over \epsilon V -1 }
\ \ ,
\\
\lambda &=&
{4\over \epsilon V b^2 }
\left[ 1 - {a^2 \over b^2 {(\epsilon V -1)}^2} \right]
\ \ ,
\end{eqnarray}
\end{mathletters}%
where $\epsilon$ is a parameter that may take the values~$\pm1$, 
\begin{equation}
  V := \sqrt{1 + 12i\eta T / b^4}
  \ \ , 
\label{V-def}
\end{equation}
and we choose the branch of the square root in (\ref{V-def}) 
so that the real part of $V$ is positive. 

It is clear from (\ref{taub-alg-sol})
and (\ref{V-def}) that all the four sign combinations for $\eta$ and
$\epsilon$ yield complex geometries that satisfy our boundary
conditions. In terms of the boundary data, the actions of these
geometries read 
\begin{equation}
  S^\eta_\epsilon (a,b;T)
  =
  {i \eta \over 3} \left[ {a^2 \over 2}
  \left( 1 - {2\over \epsilon V - 1} \right)
  - b^2 (\epsilon V - 1) \right]
  \ \ . 
  \label{taub-Setaepsilon}
\end{equation}

Following the same strategy starting from the exceptional family
(\ref{taub-class-sol-ex}) yields no solutions: regularity at $s=0$
implies $t_0=0$ and $D=0$, and positivity of the total elapsed $T$ is
then incompatible with the positivity of $b^2$ at the boundary.

None of the solution geometries we have found in this section is
Lorentzian, or indeed real with any signature.  Rather all are genuinely
complex.  Although Euclidean signature metrics have played no role in
their derivation, it may nevertheless be of interest to see what happens
when these metrics are continued to purely imaginary values of the
unimodular time~$T$.
(In particular, this may bear on the important question of whether one
can reach our saddle point metric from a Lorentzian metric without
leaving the kind of domain defined in Section 5 of 
Ref.\ \cite{yarmulke}.)
When $T$ is 
analytically continued to imaginary values as $T = -i\eta T_R$ with
$T_R>0$, it can be verified that the geometries with $\epsilon=1$
continue to Riemannian geometries that satisfy the unimodular boundary
data on~$\ballfourc$, with $T_R$ proportional to the total Riemannian
4-volume; conversely, it can be verified that these are the only
Riemannian Taub solutions on $\ballfourc$ with the unimodular boundary
data.\footnote{The singularity in (\ref{taub-alg-sol-A2}) at $V=2$ is
  a coordinate effect. A manifestly regular Riemannian form for the
  metric with $V=2$ can be found as a Riemannian section of the
  exceptional family~(\ref{taub-class-sol-ex}).}  
The continuation
sends $iS^\eta_1$ to~$-\eta I$, where $I$ is the Riemannian unimodular
action. Therefore, our complex solutions with $\epsilon=1$ are related
to solutions of the Riemannian theory by an analytic continuation in
the unimodular time: the Wick rotation is in the usual direction for
$\eta=1$ and in the unusual direction for $\eta=-1$. When the same
continuation is done to our solutions with $\epsilon=-1$, one obtains
geometries on $\ballfourc$ that satisfy the Riemannian unimodular
data, in particular with positive Riemannian volume, and
$iS^\eta_{-1}$ again continues to real values. These geometries are
however not Riemannian but complex.

\subsection{Saddle-point-estimate wave functions} 
\label{subsec:taub-b4-wavefuncs}

We turn now to the saddle-point estimate (\ref{taub-nb-estimate}) to
the no-boundary wave function~$\Psinb$, with the classical actions
$S^\eta_\epsilon$ of (\ref{taub-Setaepsilon}).  We assume throughout the
pre-exponential factors to be so slowly varying that the dominant
behavior of each term in (\ref{taub-nb-estimate})
arises from the exponentials.  We assume also that the set of saddle points
contributing to $\Psinb$ 
({\em i.e.}, 
the range of the index $k$ in (\ref{taub-nb-estimate})) 
does not depend on the values of the parameters~$T$,
$a$, and~$b$.\footnote
{In section \ref{sec:taub-cross} 
  we will have to generalize this
  assumption slightly to cover the case where two saddle points
  degenerate for certain values of $a$, $b$ and~$T$.}

It can be verified that $S^\eta_\epsilon$ satisfies the time-dependent
Hamilton-Jacobi equation with the
Hamiltonian~(\ref{taub-Hamiltonian}), as by construction it must.
Consequently, the estimate (\ref{taub-nb-estimate}) satisfies the
Schr\"odinger equation
\begin{equation}
      i {\partial \psi\over\partial T} = {\hat H} \psi
\label{schr-eq}
\end{equation}
in the approximation to which we are working.  We ask: which, if any,
of the terms in (\ref{taub-nb-estimate}) are compatible with $\Psinb$
being a wave function in the Hilbert space $\hilbert$ defined in
subsection~\ref{subsec:taub-in-unimod-gen}, evolving unitarily by some
selfadjoint 
extension of the Hamiltonian~(\ref{taub-Hamiltonian})?

Consider first the case $\eta=-1$. As the first term in
$S^{-1}_\epsilon$ has a negative imaginary part for either sign
of~$\epsilon$, the terms in (\ref{taub-nb-estimate}) with $\eta=-1$
diverge exponentially as $a\to\infty$ with fixed~$b$ (or, in the
coordinates $(u,v)$ of subsection~\ref{subsec:taub-in-unimod-gen}, as
$u\to\infty$ with fixed~$v$). These saddle points are therefore not
compatible with $\Psinb$ having finite $L^2$ norm.

Consider then the case $\eta=1$. For fixed~$T$, the imaginary part of
$S^{1}_\epsilon$ 
is bounded below.  Without further knowledge of the
pre-exponential factor, one cannot therefore exclude the wave
functions $\Psi_+ := P_+ \exp(i S^1_1)$ and $\Psi_- := P_- \exp(i
S^1_{-1})$ from being normalizable.  When $T/b^4 \ll1$, we have
\begin{mathletters}
\begin{eqnarray}
   S^1_1 &=& - {a^2 b^4 \over 18 T} + \cdots
\label{S-eta-1}
\ \ ,
\\
  S^1_{-1} &=& { i (a^2 + 2b^2) \over 3} + \cdots
\label{S-eta--1}
\ \ .
\end{eqnarray}
\end{mathletters}%
In this limit, the wave function $\Psi_+$ is therefore rapidly oscillating,
while~$\Psi_-$, on the other hand, is
exponentially suppressed for large $a$ or~$b$.
One can also see that in this limit, the
saddle point metric 
with $\epsilon=1=\eta$ is
(except near $s=0$)
close to a Lorentzian 
\tnds\ universe with $m=0$. 
We note that a similar conclusion 
was reached within the conventional
Einstein theory in Ref.\ \cite{jlr}.  

To address the unitarity of the evolution, we note that for fixed $a$
and~$b$, $S^1_\epsilon$ has the large $T$ expansion
\begin{equation}
S^1_\epsilon =
2 \epsilon e^{-i\pi/4} \sqrt{T/3}
+ \case{1}{6} i (a^2 + 2b^2)
+ O\left(T^{-1/2}\right)
\ \ .
\label{taub-S1eps-as-out}
\end{equation}
The contribution to the inner product $(\Psi_+,\Psi_+)$ from any
compact region in the configuration space is therefore exponentially
increasing in~$T$.  This means that $\Psi_+$ cannot evolve by any
selfadjoint extension of the Hamiltonian~(\ref{taub-Hamiltonian}):
probability is being injected into the configuration space either from
the finite boundaries or from infinity.
As we do not have an estimate for the saddle-point prefactor, we shall
not attempt to discuss exactly where the probability is entering the
configuration space.
However, if the saddle-point form is good for $T/b^4\gg1$, equation
(\ref{taub-S1eps-as-out}) shows that the probability is flowing in the
direction of isotropic {\em expansion\/},
as measured by ${\rm Imag}(\Psi^*\nabla\Psi)$;
and for $T/b^4\ll1$, equation (\ref{S-eta-1}) 
exhibits a similar direction of flow.
This suggests that the flux
is coming from somewhere on the finite boundary. 
The emerging picture
is perhaps compatible with the tunneling proposals 
advocated by Linde 
\cite{linde-zetf,linde-ncim,linde-repprog,linde-mezh,linde2-mezh} 
and Vilenkin 
\cite{vile1,vile2a,vile2b,vile3,vile-erice,vile-discord}. 
Notice, however, that the latter proposals are couched in terms of the
non-unimodular theory, where the relevant boundary is (for the Taub
model) only one-dimensional.  For us, it is two-dimensional (the extra
dimension being parameterized by~$T$), and there can be no simple
correspondence between boundary conditions formulated in the two
frameworks. 

In this connection notice that, precisely because the unimodular $\psi$
depends on a parameter time $T$, the early-time behavior of the universe
shows up much more directly in $\psi$ than it would in a non-unimodular
formulation, and one can ask whether $\psi$ is concentrated for $T=0$ at
universes of zero size, as the no-boundary picture would seem to
require.  
For the wave function $\Psi_+$, equation
(\ref{S-eta-1}) does appear to describe a state for which the probability is
concentrated near configurations of small 3-volume (in the sense that
the rapid oscillations cause $\Psi_+$ to vanish uniformly in the
distributional sense as $T\to{0}$, in any region of compact support
disjoint from the boundary $uv=0$).\footnote%
{More precisely, $\Psi_+$ seems to describe a violent explosion starting
from zero volume, analogously to how a nonrelativistic particle
initially at the origin at $t=0$ is at any later moment uniformly
distributed throughout space with a wave function $\propto\exp(ix^2/t)$.}
For the wave function $\Psi_{-}$,
equation (\ref{S-eta--1}) describes a universe that at birth is of Planckian
dimensions in all directions, which for quantum gravitational purposes
is probably as close as one would care to get to strictly zero size.

With $\epsilon=-1$, we see from (\ref{taub-S1eps-as-out}) that the
contribution to the inner product $(\Psi_-,\Psi_-)$ from any compact
region in the configuration space is exponentially decreasing in~$T$.
One might see this as evidence for a different type of non-unitary evolution
of~$\Psi_-$, with probability now flowing out of the 
configuration space.
Perhaps, however, one cannot exclude that $\Psi_-$ might
still approximate a unitarily evolving wave function in some relevant
sense: the probability would just be spreading out sufficiently fast to
give an exponential suppression at late times in any fixed compact region
of the configuration space.  Resolving this question would seem to require,
at a minimum, a better understanding of the prefactor in our saddle-point
estimate.

In summary, the wave functions with $\eta=-1$ cannot represent states
in the Hilbert space~$\hilbert$. The wave function with
$\eta=\epsilon=1$ may represent a state in~$\hilbert$, but it
has an exponentially growing norm and thus cannot evolve unitarily.
The wave function with $\eta=1$ and $\epsilon=-1$ may represent a
state in the Hilbert space, and, at the level of our semiclassical
estimate, this state may evolve unitarily.  It is therefore the only
one that might be compatible, in a Lorentzian histories framework,
with a universe consisting of a single macroscopic component with a
single moment of birth.

\section{Taub no-boundary saddle points for cross-cap topology}
\label{sec:taub-cross}

In this section we discuss the Taub
saddle points and wave functions when the (truncated)
4-manifold is the closed 4-dimensional cross-cap,
$\crosscapfourc := \RPfour \setminus B^4 \simeq \RPfour \#
\ballfourc$, where $B^4$ is the open 4-dimensional
ball.\footnote{This manifold was suggested to us by John Friedman.}
In subsection \ref{subsec:taub-cross-geoms} we find the 
saddle-point geometries, and in subsection
\ref{subsec:taub-cross-wavefuncs} we discuss the saddle-point
estimates to the wave function.

\subsection{Saddle-point geometries} 
\label{subsec:taub-cross-geoms}

It is useful to view $\crosscapfourc$ as a quotient space. To this
end, let $\Mtilde := [-1,1] \times S^3$. $\Mtilde$~is a compact
orientable manifold with boundary $S^3 \cup S^3$. Consider the map $J:
\Mtilde\to\Mtilde$; $(s,\bfx) \mapsto (-s,P\bfx)$, where $P: \bfx
\mapsto P\bfx$ is the antipodal map on~$S^3$. $J$~is an involution
with a free and properly discontinuous action, and the quotient space
$\Mtilde/J$ is diffeomorphic to~$\crosscapfourc$.  One can also visualize
$\crosscapfourc$ as built by closing off the upper half $s>0$
of $\Mtilde$ 
by attaching $s=0$ to an~$\RPthree$. 
Clearly, $\crosscapfourc$~is
a nonorientable compact manifold with boundary~$S^3$.

We need first the general complex Taub geometry on~$\crosscapfourc$.
To begin, we recall from 
section \ref{sec:taub-b4}
that the complex Taub geometries on $\Mtilde$ are
obtained from (\ref{taub-class-sol}) or
(\ref{taub-class-sol-ex}) by
letting the parameters take complex values and writing the coordinate
$t$ as a complex-valued function~$t(s)$, with $dt/ds\ne0$. 
We take the boundaries of $\Mtilde$ to be at $s=\pm1$. 
On these geometries, we
realize $J$ as the map $(s,\bfx) \mapsto (-s,P\bfx)$, where $P$ acts
on $\sutwo\simeq S^3$ as multiplication by ${\rm{diag}}(-1,-1)$ in the
defining matrix representation. In order that $J$ be an isometry, it
is necessary that $db/ds$ and $da/ds$ vanish at $s=0$. This
excludes~(\ref{taub-class-sol-ex}), and shows that
(\ref{taub-class-sol}) is only possible with $t(0)=0$ and $B=0$.  The
condition that $J$ be an isometry 
implies, finally, $t(s)=-t(-s)$.
Quotienting this geometry on $\Mtilde$ with respect to $J$ now
gives the desired general complex geometry on~$\crosscapfourc$.

Next, we need to match the complex geometry on $\crosscapfourc$ to the
unimodular boundary data. 
{}From 
$t(1)=-t(-1)=:\tstar$, 
we have
\begin{equation}
  T = \int_0^{\tstar} Nab^2 \, dt
  = \tstar \left( A^2 +
  {\tstar^2 \over 3 A^2} \right)
\ \ ,
\label{taub-cross-Ttotal}
\end{equation}
and the action is evaluated from (\ref{taub-uaction}) to be 
\begin{equation}
  S =
  {\lambda \tstar \over 6}
  \left( -5 A^2 - {\tstar^2 \over A^2}
  + {8 A^6 \over A^4 + \tstar^2 } \right)
  + {4 \tstar^3 \over 3 (A^4 + \tstar^2)}
\ \ .
\label{taub-cross-intaction}
\end{equation}
The values of $\lambda$, $A^2$, and
$\tstar$ can now be found in terms of $T$ and the final scale factors
from~(\ref{taub-class-sol-b}), (\ref{taub-class-sol-aabb}),
and~(\ref{taub-cross-Ttotal}). The general solution to this system of
algebraic equations is
\begin{mathletters}
\label{taub-cross-algsol} 
\begin{eqnarray}
\tstar &=& y b^2/z
\ \ ,
\\
A^2 &=&
b^2/z
\ \ ,
\label{taub-cross-Atwosol}
\\
\lambda &=&
{ z \left[ (a^2/b^2) z^2 + 4z -8 \right]
\over
b^2 \left( z^2 + 4z -8 \right) }
\ \ ,
\label{taub-cross-lambdasol}
\end{eqnarray}
\end{mathletters}%
where
\begin{mathletters}
\label{zy-def}
\begin{equation}
z = 1 + y^2
\ \ ,
\label{z-def}
\end{equation}
and $y$ is any solution to
\begin{equation}
{3T \over b^4 } =
{y (3 + y^2) \over {(1 + y^2)}^2 }
\ \ .
\label{y-def}
\end{equation}
\end{mathletters}%
Every root of (\ref{y-def}) does, with one exception, yield a
saddle-point geometry on $\crosscapfourc$ with our boundary
conditions. The exception occurs when $z=2 \left( \sqrt{3} -1
\right)$, in which case the denominator of
(\ref{taub-cross-lambdasol}) vanishes.

With (\ref{taub-cross-algsol}) and~(\ref{zy-def}), the action
(\ref{taub-cross-intaction}) can be written as
\begin{equation}
  S
  =
  {Tz \over 2 b^2 (2+z)} \left( 4 - {a^2 z \over b^2} \right)
  \ \ .
\label{taub-cross-actionz}
\end{equation}
As (\ref{taub-cross-actionz}) depends on $y$ only through~$z$, the
distinct saddle-point values of the action are obtained by finding the
distinct values of $z$ from~(\ref{zy-def}). This is equivalent to
solving for $z$ the quartic
\begin{equation}
{\left( 3T \over b^4 \right)}^2 =
{(z-1){(z+2)}^2 \over z^4}
\ \ .
\label{zbare-def}
\end{equation}
It is easily verified that the action~(\ref{taub-cross-actionz}), with
$z$ given by any root of (\ref{zbare-def}) that is a smooth function
of~$T/b^4$, satisfies the Hamilton-Jacobi equation.

The roots of (\ref{zbare-def}) depend on the quantity $\alpha :=
3T/b^4>0$. Let
\begin{mathletters}
\begin{eqnarray}
\alpha_c &:=&
2^{-5/2} 3^{3/4} \left( \sqrt{3} + 1 \right)
\ \ ,
\\
z_c &:=&
2 \left( \sqrt{3} -1 \right)
\ \ .
\end{eqnarray}
\end{mathletters}%
For $\alpha<\alpha_c$, there are two distinct real roots, denoted by
$z_+$ and~$z_-$, satisfying $1< z_- < z_c < z_+$.  For
$\alpha=\alpha_c$, these two roots merge at~$z_c$: this is the special
case in which (\ref{taub-cross-lambdasol}) becomes singular.  For
$\alpha>\alpha_c$, these two roots become a complex conjugate pair,
denoted by $(z_1,z_2)$, where $z_1$ is in the first quadrant and $z_2
= {\bar z}_1$. The two remaining roots are always a complex conjugate
pair, denoted by $(z_3, z_4)$, where $z_3$ is in the second quadrant
and $z_4 = {\bar z}_3$. The only instance in which a
root is not a smooth function of $\alpha$ occurs in the transition of
the pair $(z_+,z_-)$ into the pair $(z_1,z_2)$ at $\alpha=\alpha_c$.

The roots $z_+$ and~$z_-$ of (\ref{zbare-def}) give Lorentzian saddle-point
geometries on $\crosscapfourc$. These geometries
have the peculiarity that they are not time-orientable, and they
therefore would fall into the framework of time-nonorientable
cobordisms \cite{non-t-orient}, if they were regarded as being in the
domain of the Lorentzian path
integral.\footnote%
{For a discussion of quantum field theory on time-nonorientable
  spacetimes, see Ref.\ \cite{fried-higu}.  
  Because of their time-non-orientability, we suspect that these
  metrics are not valid saddle points for a path integral that is
  originally taken over almost everywhere Lorentzian metrics with a 
  well-defined causal structure.  
  See section \ref{sec:discussion} for some
  further thoughts on how one might in principle recognize the valid
  saddle points.} 
All the other saddle-point geometries are complex.

We denote the action (\ref{taub-cross-actionz}) evaluated at the
respective roots by $S^+$, $S^-$, $S^1$, $S^2$, $S^3$, and~$S^4$. For
later use, we note that the roots have for
$\alpha\ll1$ and
$\alpha\gg1$ the respective expansions
\begin{mathletters}
\begin{eqnarray}
z_+ &=& \alpha^{-2} + 3 + O (\alpha^2)
\\
z_- &=& 1 + \case{1}{9} \alpha^2 + O (\alpha^4)
\\
z_3 = {\bar z}_4 &=& -2 + {4i\alpha\over\sqrt{3}}
\left[ 1 - {10i\alpha \over 3 \sqrt{3} }
+ O (\alpha^2 ) \right]
\ \ ,
\end{eqnarray}
\end{mathletters}%
and
\begin{mathletters}
\label{zetas-large-alpha}
\begin{eqnarray}
z_1 = {\bar z}_2 &=& e^{i\pi/4} \sqrt{{2/\alpha}}
\left[ 1 + O ( \alpha^{-1} ) \right]
\ \ ,
\\
z_3 = {\bar z}_4 &=& e^{3i\pi/4} \sqrt{{2/\alpha}}
\left[ 1 + O ( \alpha^{-1} ) \right]
\ \ .
\end{eqnarray}
\end{mathletters}%

\subsection{Saddle-point-estimate wave functions} 
\label{subsec:taub-cross-wavefuncs}

We are now ready to consider the saddle-point estimate
(\ref{taub-nb-estimate}) to the no-boundary wave function. As in
section~\ref{sec:taub-b4}, we assume throughout that the pre-exponential
factors are not important, and that the set of saddle points contributing
to $\Psi$ does not change in ranges of $(T,a,b)$ where the corresponding
saddle point geometries vary continuously.
As before, we write $\Psi_+$, $\Psi_-$, $\Psi_1 $, 
and so on, 
for the saddle
point wave functions corresponding to $S^+$, $S^-$, $S^1$, etc.  Notice,
however, that while $\Psi_3$ and $\Psi_4$ are defined on all of
configuration space, the pair $\Psi_\pm$ (respectively $\Psi_1$, $\Psi_2$)
is defined only for $\alpha<\alpha_c$  (respectively $\alpha>\alpha_c$).

It can be shown that the second term in $S^1$ and $S^4$ has a negative
imaginary part.  
{}From 
the limit $a\to\infty$ with fixed $b$ and~$T$,
it is therefore seen that $\Psi_1$ and $\Psi_4$ 
are not normalizable.  

The imaginary part of~$S^3$, on the other hand, is positive, and the
corresponding term $\Psi_3$ in (\ref{taub-nb-estimate}) may represent a
state in~$\hilbert$.  For $\alpha\ll1$ with $a\approx b$, we have
\begin{equation}
  S^3 = {i(a^2 + 2 b^2) \over 2\sqrt{3}} + \cdots 
\ \ .
\end{equation}
This again describes a Planck sized universe at $T=0$.
The contribution to the inner product $(\Psi_3,\Psi_3)$ from any
compact region in the configuration space is exponentially decreasing
in~$T$ (see equation (\ref{taub-cross-S3largeT}) below).  
As in section~\ref{sec:taub-b4}, we regard this as consistent
with unitary evolution, although it also might signify a loss of
probability through the finite boundary.  

The only remaining possibility in (\ref{taub-nb-estimate}) is a wave
function that coincides with $\Psi_2=P_2 \exp(iS^2)$ for
$\alpha>\alpha_c$, and with some combination of $\Psi_+$ and $\Psi_-$
for $\alpha<\alpha_c$. We denote this wave function by~$\Psi_0$. The
curve $\alpha=\alpha_c$ in the configuration space is analogous to a
turning point in a constant-energy WKB approximation, and a
saddle-point estimate to $\Psi_0$ would not be expected to be good
near this curve.  However, beginning as $\Psi_2$ for
$\alpha>\alpha_c$, $\Psi_0$ presumably resumes a saddle-point form for
$\alpha<\alpha_c$, now as a linear combination of $\Psi_+$ and
$\Psi_-$.  Since the imaginary part of $S^2$ can be shown to be
bounded below, and $S^\pm$ are purely real, $\Psi_0$~may thus be
normalizable.

When $\alpha<\alpha_c$, the two terms in $\Psi_0$ have 
each an immediate semiclassical interpretation, as 
they each come from a Lorentzian
universe (\ref{taub-class-sol}) with $B=0$.  The two parameters in
this family are $A$ and~$\lambda$. The reason why {\em two\/} such
solutions exist is that there are two choices for $A$ and~$\lambda$,
obtained from (\ref{taub-cross-Atwosol}) and
(\ref{taub-cross-lambdasol}) with respectively~$z=z_\pm$, that make a
spacetime in this family pass through a prescribed point in the
configuration space at a prescribed value of~$T$.  For
$\alpha=3T/b^4\ll1$ with $a\approx b$, we have
\begin{mathletters}
\label{Spm-as-small-alpha}
\begin{eqnarray}
S^+ &=& - {a^2 b^4 \over 18 T} + \cdots
\ \ ,
\\
S^- &=&
{T \over 6 b^2 } \left( 4 - {a^2 \over b^2} \right)
+ \cdots
\ \ .
\end{eqnarray}
\end{mathletters}%
{}From (\ref{Spm-as-small-alpha})
we also see that at $T=0$, $\Psi_+$ behaves
similarly to $\Psi_+$ in the previous section, as if the universe had
exploded from zero size.  On the other hand, $\Psi_-$ is quite
different from its earlier namesake, looking like a zero momentum
state spread out  over all of configuration space.

Consider, finally, whether the norms of $\Psi_0$ and~$\Psi_3$ can be
independent of~$T$.  For fixed $a$ and~$b$, (\ref{zetas-large-alpha})
gives the large $T$ expansion
\begin{mathletters}
\begin{eqnarray}
 S^2 &=&
 e^{-i\pi/4} \sqrt{{2T/3}} + {i (a^2 + 2b^2) \over 6 }
 + O \! \left( T^{-1/2} \right)
\ \ ,
\label{taub-cross-S2largeT}
\\
 S^3 &=&
 e^{3i\pi/4} \sqrt{{2T/3}} + {i (a^2 + 2b^2) \over 6 }
 + O \! \left( T^{-1/2} \right)
\ \ .
\label{taub-cross-S3largeT}
\end{eqnarray}
\end{mathletters}%
It follows that $\Psi_3$ dies out with time at any fixed $a$, $b$, 
whence its evolution can be unitary.
On the other hand,
the contribution to the inner product $(\Psi_0,\Psi_0)$ from any
compact region in the configuration space is 
seen to be exponentially
increasing in~$T$, and $\Psi_0$ cannot evolve unitarily. As in
section~\ref{sec:taub-b4}, it is difficult to ascertain where the
probability is entering the configuration space, but the semiclassical
discussion given above for $\alpha\ll1$ suggests that the flux may
be coming from somewhere on the finite boundary, as before.

In summary, 
the qualitative results 
with the 4-manifold $\crosscapfourc$
are very similar to 
those obtained 
with the 4-manifold $\ballfourc$ 
in section~\ref{sec:taub-b4}.  
Of the four saddle points, 
two lead to non-square-integrable wave functions,
analogously to the case $\eta=-1$ earlier.  
The normalizable cases here are those of $\Psi_3$ and~$\Psi_0$.  
The wave function $\Psi_3$ here is analogous to $\Psi_-$ there: 
it may be normalizable, 
it evolves consistently with unitarity, 
it is nowhere rapidly oscillating, 
and it describes a universe 
that has Planckian size at $T=0$.  
Similarly, 
the wave function $\Psi_0$ here 
is analogous to $\Psi_+$ there: 
it may be normalizable, 
it has a WKB form 
expressing a classically evolving universe in a suitable limit, 
but its norm cannot be independent of~$T$.  
Its behavior at $T=0$ differs from that of the earlier~$\Psi_+$, 
however, 
in a way related to 
the degeneracy of $z_1$ and $z_2$ at $\alpha=\alpha_c$.

\section{Friedmann truncation of the Taub 
saddle points and wave functions}
\label{sec:fried-restr}

The Taub saddle-point metrics found in sections \ref{sec:taub-b4} and
\ref{sec:taub-cross} clearly specialize to saddle-points of the
Friedmann model by setting $a=b$, and the corresponding actions are the
restrictions to $a=b$ of those found earlier.

For most of the saddle points, the discussion within the Friedmann
model proceeds in parallel
with that in the Taub model.\footnote%
{The results reported for the Friedmann model with the 4-manifold
  $\ballfourc$ in Ref.\ \cite{dls-waterloo} only considered the saddle
  points with $\epsilon=1$, inadvertently excluding the saddle points
  with $\epsilon=-1$.}
In particular, with the 4-manifold~$\ballfourc$, the saddle point
with $\eta=\epsilon=1$ yields an exponentially growing probability
flux, and this flux must now enter the configuration space at the
boundary $a=0$. With the 4-manifold~$\crosscapfourc$, a similar
argument can be made for the wave function~$\Psi_0$.

The only qualitative difference between the Taub analysis and the
Friedmann analysis occurs for the saddle point with $\eta=-1$ and
$\epsilon=1$ on~$\ballfourc$, and for the saddle point with $z_1$
on~$\crosscapfourc$.  In the two Taub models, the corresponding wave
functions (call them $\Psi^-_+$ and $\Psi^-_1$) were seen not to be
normalizable.  In the Friedmann situation, however, the imaginary part
of the action turns out to be bounded below and the $a=b$ restrictions
of these wave functions are in fact square integrable.
(With~$\crosscapfourc$, $\Psi^-_1$ covers only part of the
configuration space.  However, when $\alpha<\alpha_c$, it becomes a
linear combination of terms arising from~$S^\pm$ , so the full wave
function $\Psi^-_0$ is also normalizable.)  In the Friedmann
restriction, therefore, these saddle points are both compatible with
normalizability.  Moreover, for fixed~$a$,
these 
wave functions are exponentially decreasing in~$T$,
and are even compatible with unitary evolution.

The drastic qualitative change in these two saddle-point contributions
upon passing from the Friedmann model to the Taub model suggests that
the ``good" behavior of these saddle points in the Friedmann model
should be seen as an artifact of the isotropic truncation. We shall
return to this question in section~\ref{sec:discussion}.

\section{Bianchi type~I}
\label{sec:bianchiI}

In this section we discuss the unimodular no-boundary saddle points
and wave functions in a Bianchi type~I minisuperspace model.  We take
the spatial topology to be~$T^3$, and the (truncated) no-boundary
4-manifold to be 
$\BbbD^2\times{T^2}$, where $\BbbD^2$ is the closed disk.  
We set up the unimodular quantum theory in
subsection~\ref{subsec:ham-bianchiI}. The no-boundary saddle points
and wave functions are analyzed in
subsection~\ref{subsec:nb-bianchiI}.

\subsection{Unimodular quantization of Bianchi type~I}
\label{subsec:ham-bianchiI}

The local spatial homogeneity of Bianchi type I is compatible with ten
distinct closed spatial topologies \cite{wolf}. The number of
minisuperspace degrees of freedom depends on the spatial topology
\cite{ellis,louko-ruback,giu-lou-thetas,%
hoso+1,waelbroeck,hoso+2,hoso+3,kodama-homog},
and the spatial topology also determines the group of large spatial
diffeomorphisms that can be incorporated as gauge invariances of the
model \cite{louko-ruback,giu-lou-thetas}.  The topology also
determines the possible ways of compactifying the manifold toward the
past to obtain a manifold of no boundary
type\cite{louko-ruback,giu-lou-thetas}.

We shall here focus 
on a Bianchi type I model with an additional
discrete symmetry group reminiscent of the additional $\uone$ symmetry
that distinguishes the Taub class of metrics within Bianchi
type~IX\null.  The results obtained for the conventional Einstein
theory in Refs.\ \cite{louko-ruback,giu-lou-thetas} suggest that this
specialized model should faithfully reflect the general Bianchi type I
situation regarding the normalizability and unitary evolution of the
wave function.

The metric of our Bianchi type I model reads
\begin{equation}
ds^2 = \rho^2 \left[
- N^2 dt^2 +  a^2 dx^2
+ b^2 \! \left( dy^2 + dz^2 \right)
\right]
\ \ ,
\label{bI-metric}
\end{equation}
where $a$, $b$, and $N$ are functions of~$t$, and the overall factor
$\rho^2 := G/(2\pi^2)$ has been introduced for numerical
convenience. 
In this subsection we take $a^2$, $b^2$, and
$N^2$ to be positive, so that the metric is Lorentzian. 
The identifications made on the spatial
coordinates are $(x,y,z) \sim (x+2\pi,y,z) \sim (x,y+2\pi,z) \sim
(x,y,z+2\pi)$, and the spatial topology is thus~$T^3$.  The metric
(\ref{bI-metric}) is obtained from the most general Bianchi type I
metric with $T^3$ spatial topology by imposing the extra symmetry
${\Bbb Z}_2\times D_8$ where the 
8-element 
dihedral group $D_8$ is the
symmetry group of the square.  This is equivalent to demanding that
the spatial metric have three {\em orthogonal\/} closed geodesics, and
that two of these geodesics have equal length.

To derive the solutions of the conventional Einstein theory and the
unimodular theory, we proceed as in section~\ref{sec:taub-in-unimod}.
Inserting the metric (\ref{bI-metric}) into the action-integral
(\ref{S-grav}) with bare cosmological constant~$\Lambda$, and
introducing the proper time parameter $\tau$ by $d\tau = Ndt$, we
obtain the minisuperspace action
\begin{equation}
  S = \case{1}{2}
      \int d\tau
      \left[
      - a \left({db\over d\tau}\right)^2 
      - 2b {da\over d\tau} {db\over d\tau} 
      - \lambda a b^2 
      \right]
\ \ ,
\label{bI-conv-action}
\end{equation}
where $\lambda := \rho^2 \Lambda$. This action reproduces the full
Bianchi type I Einstein equations with a cosmological constant under
variations that fix the initial and final values of the scale factors
but not those of~$\tau$. For $\lambda\ne0$, the general Lorentzian
solution can be written in the gauge $Na=1$ as
\begin{mathletters}
\label{BI-class-sol}
\begin{eqnarray}
b &=& At
\ \ ,
\label{BI-class-sol-b}
\\
a^2b &=& 
\case{1}{3} \lambda A t^3 + E 
\ \ ,
\label{BI-class-sol-aab}
\\
N &=& 1/a
\ \ ,
\end{eqnarray}
\end{mathletters}%
where $A\ne0$ and $E$ are constants. For $\lambda=0$, the solutions
not obtained from (\ref{BI-class-sol}) with $\lambda=0$ 
can be put in
the form
\begin{mathletters}
\label{BI-class-sol-ex}
\begin{eqnarray}
b &=& B 
\ \ ,
\\
a^2b &=& Dt + E
\ \ ,
\\
N &=& 1/a
\ \ ,
\end{eqnarray}
\end{mathletters}%
where $B\ne0$, $D$, and $E$ are constants, $D$~and $E$ not
both equal to zero. 

In order to put the action-integral in a form convenient for the
unimodular theory, we introduce the parameter time $T$ by $dT = ab^2
d\tau$.  As before, we also simplify the action, without loss of
generality in the unimodular theory, by setting the bare cosmological
constant to zero.  The integral (\ref{bI-conv-action}) then takes the
form
\begin{equation}
  S = - \case{1}{2} \int dT \, a b^2 (a b'^2 + 2b a' b')
\ \ , 
\label{bI-uaction}
\end{equation}
where the prime denotes derivative with respect to~$T$. The
4-volume bounded by the hypersurfaces $T=T_1$ and $T=T_2$, with
$T_1<T_2$, is $8\pi^3 \rho^4 (T_2-T_1)$, and fixing the 4-volume in
the variation of (\ref{bI-uaction}) is therefore equivalent to fixing
the initial and final values of~$T$. The unimodular variational
equations clearly reproduce the Einstein equations, with the
cosmological constant now emerging as the integration constant
proportional to the unimodular energy.

{}From~(\ref{bI-uaction}), the unimodular Hamiltonian operator is
\begin{equation}
{\hat H} := 6 {\partial^2 \over \partial u \partial v}
\ \ ,
\label{bI-Hamiltonian}
\end{equation}
where the coordinates $(u,v)$ are defined by~(\ref{uv-def}), and we
have adopted the Laplace-Beltrami factor ordering, as in
section~\ref{sec:taub-in-unimod}. The matching inner product is
again~(\ref{ip}).  As ${\hat H}$ is symmetric and real, it has
selfadjoint 
extensions by von Neumann's theorem.

\subsection{No-boundary saddle points and wave functions}
\label{subsec:nb-bianchiI}

The general complex solution with our Bianchi I symmetries is obtained
from the Lorentzian solutions (\ref{BI-class-sol}) and
(\ref{BI-class-sol-ex}) by extending the parameters to complex values
and making $t$ a complex-valued function $t(s)$ of a real-valued time
coordinate~$s$. We may assume $dt/ds\ne0$. The condition that the
solution be defined on $\BbbD^2\times{T^2}$ means that $s$ must be
interpretable as the radial coordinate of polar coordinates
on~$\BbbD^2$.  Taking $s\in[0,1]$ as in
subsection~\ref{subsec:taub-b4-geoms}, with $s=0$ occurring at the
coordinate singularity, it is then necessary that $a$ vanish at $s=0$
but $b$ remain nonzero there. It is straightforward to show from
(\ref{BI-class-sol}) and (\ref{BI-class-sol-ex}) that the general
complex solution with this property can be written as
\begin{mathletters}
\label{BI-class-sol-zero}
\begin{eqnarray}
b &=& 
B \left( 1 + \lambda \Atilde t \right) 
\ \ ,
\label{BI-class-sol-zero-b}
\\
a^2b &=& t (B/\Atilde)
\left( 
1 
+ \lambda \Atilde t
+ \case{1}{3} \lambda^2  \Atilde^2 t^2 
\right) 
\ \ ,
\label{BI-class-sol-zero-aab}
\\
N &=& 1/a
\ \ ,
\end{eqnarray}
\end{mathletters}%
where $B$ and $\Atilde$ are nonvanishing complex constants, and we
have chosen $t(0)=0$. The absence of a conical singularity at $s=0$
requires $N^{-1} da/dt \to i\eta$ as $s\to0$, where $\eta$ is a
parameter that takes the values~$\pm1$: this implies $\Atilde =
-\casehalf i \eta$.  The metric then defines a solution on $\BbbD^2
\times T^2$ in the sense we seek.

The total elapsed $T$ is
\begin{equation}
  T = \int_0^{\tstar} Nab^2 \, dt
  = \tstar B^2 
\left( 
1 
-\casehalf i\eta \lambda \tstar
- \case{1}{12} \lambda^2 \tstar^2 
\right) 
\ \ ,
\label{BI-totalT}
\end{equation}
where we have written $t(1)=:\tstar$. 
Solving for~$B$, $\lambda$, and
$\tstar$ in terms of $T$ and the boundary values of the scale factors,
we obtain
\begin{mathletters}
\label{BI-Blt}
\begin{eqnarray}
B &=&
{2 i\eta T \over a^2b}
\ \ ,
\\
\lambda &=&
{8T \over 3 a^4 b^2}
\left[
\left( a^2 b^2 \over 2T \right)^3
+ i \eta
\right]
\ \ ,
\\
\tstar &=&
{ 3 a^4 b^2
\over
{\displaystyle{
4 T
\left[
\left( a^2 b^2 \over 2T \right)^2
+ {i \eta a^2 b^2 \over 2 T}
- 1
\right]
}}
}
\ \ .
\end{eqnarray}
\end{mathletters}%
As discussed in subsection~\ref{subsec:taub-b4-geoms}, 
the action contains the integral term (\ref{taub-uaction}) 
as well as a boundary term from $s=0$, 
and for the metric (\ref{bI-metric}) 
the boundary term reads\footnote%
{The boundary term contributes here because we are essentially in a
2-dimensional situation.  Its presence marks a genuine difference
between the point of view that the spacetime manifold is a cobordism with
empty initial boundary, and the point of view that it has an initial
boundary of zero spatial volume.  
In the Taub models we considered, this
distinction was effectively moot because the analogous boundary term did
not contribute.  
Similarly, the assumption that the saddle point
metric must be smooth is also playing an important role here, in
contrast to the Taub case.
The
boundary term is somewhat reminiscent of
the pure imaginary topological contribution 
to the {\em Lorentzian\/} action-integral  
found in Ref.\ \cite{yarmulke}. }
$-\casehalf N^{-1} d(ab^2)/dt$
\cite{jjhlou3,laf-stwo,lou-annphys,lou-canoniz}. The value of the
integral term is $-\casehalf \lambda T$, and that of the boundary term
is $-\casehalf i \eta B^2$. In terms of the boundary data, the action
reads
\begin{equation}
  S^\eta (a,b;T)
  =
  - {a^2 b^4 \over 6 T}
  + {2 i\eta T^2 \over 3 a^4 b^2}
\ \ .
\end{equation}
It is easily verified that this action satisfies the Hamilton-Jacobi
equation. 

The solution geometries are genuinely complex,  analogously to those found
for the Taub model with $\Reals^4$ (untruncated) topology. 
As before, we have not tried to analyze directly which, if any, of
the saddle points can be reached from an almost Lorentzian metric on the
same manifold by an admissible deformation.  We can, however, glean some
indirect evidence on this by considering the Wick rotation to
the 
{\em Riemannian\/} case.  
When $T$ is analytically continued to
imaginary values as $T = -i\eta T_R$ with $T_R>0$, the geometries
continue to Riemannian geometries that satisfy the unimodular boundary
data on~$\BbbD^2 \times T^2$, with $T_R$ proportional to the total
Riemannian 4-volume; conversely, these are the only Riemannian
solutions of our Bianchi type I model on $\BbbD^2 \times T^2$ with the
unimodular boundary data. The continuation sends $iS^\eta$ to~$-\eta
I$, where $I$ is the Riemannian unimodular action.  The situation is
thus as for $\epsilon=1$ in section~\ref{sec:taub-b4}: the complex
spacetimes with the Lorentzian boundary conditions are related to
solutions of the Riemannian theory by an analytic continuation in the
unimodular time, with a Wick rotation in the usual direction for
$\eta=1$ and in the unusual direction for $\eta=-1$.

We can now turn to the saddle-point estimate~(\ref{taub-nb-estimate}).
For fixed~$T$, the wave function with $\eta=-1$ diverges exponentially
as $a^2b\to0$, and consequently cannot be square integrable.\footnote
{Unlike for the unnormalizable wave functions in the Taub cases, the
divergence here is for small universes rather than large ones.  In that
sense, the argument for dismissing this saddle point is perhaps less
compelling than before, because an ultraviolet cutoff on $a$ and $b$
would render the $\eta=-1$ wave function compatible with normalizability.}
The wave
function with $\eta=+1$, on the other hand, is compatible with
normalizability.  Moreover, this wave function decays exponentially as
$T\to\infty$ at fixed $a$ and~$b$.

As explained in section~\ref{sec:taub-in-unimod}, we officially regard
such behavior as consistent with unitary evolution.  In this case,
moreover, it appears  plausible that probability actually is
flowing toward infinity, rather than escaping through the finite
boundary.  Indeed, if we limit ourselves to values of $a$,$b$ $\ge{1}$
(meaning that none of the dimensions of 3-space has sub-Planckian
scale), then it is easy to see that the estimate $\Psi=O(1)e^{iS}$
implies that the probability for the 3-volume $V=ab^2$ to be less than
$\sqrt{T}$ is exponentially small in $T^2/V^4$.  Here it is helpful to
rewrite $S^\eta$ in the form
\begin{equation}
  iS^\eta (a,b;T)
  =
  - {iuv \over 6 T}
  - {2 \eta T^2 \over 3 u^2}
\ \ ,
\label{S-BI}
\end{equation}
where $u=a^2b$ and $v=b^3$ are the ``light cone coordinates'' introduced
earlier, for which $V=\sqrt{uv}$.  Thus, the universe inevitably expands
as $T$ increases.  

Moreover, if the universe
expands enough so that $a^2b^2$ becomes $\gg{T}$, then
it enters a regime in which (for both $\eta=+1$ and $\eta=-1$) $\Psi$
oscillates rapidly, with the corresponding WKB trajectories forming
(as (\ref{S-BI}) shows) a two-parameter family of classical Lorentzian
solutions that are locally isometric to de~Sitter, expanding
exponentially in the cosmological time, with the ratio of the scale
factors remaining constant.  (One parameter of the family is the
cosmological constant, and the other one is the ratio of the scale
factors.)  Indeed, the saddle point metric (\ref{BI-class-sol-zero}),
(\ref{BI-Blt}) in this regime is itself 
(with our choice of ``complex gauge'' for it)
very close to being Lorentzian
at late times, and therfore close to some specific Lorentzian solution
$\hat{g}$ of the classical Einstein equations.  
This indicates that
the major contribution to (\ref{taub-nb-integral}) for $a^2b^2\gg{T}$
comes from 4-geometries that at late times are close to $\hat{g}$, and
therefore are behaving essentially classically.

The behavior of our saddle point estimate for $T\to{0}$ is also
suggestive.  In this limit, (\ref{S-BI}) shows, as before, that the
distributional support of $\Psi$ shrinks down on $uv=0$, describing the
explosive birth of a universe at zero 3-volume $ab^2$ (although $a$ and
$b$ need not vanish separately).

In summary, only the saddle point with $\eta=1$ yields a square
integrable wave function.  This wave function is also compatible with
unitary evolution at the level of our saddle-point estimate, and it can
be construed as describing a universe that begins at 0 volume and
ultimately enters a regime of classical isotropic expansion at late
times.

In concluding this section, we mention that it would be straightforward
to analyze also the Bianchi type I analog of the 
cross-cap manifold we considered in section~\ref{sec:taub-cross}, 
namely the 4-manifold that is the
product of $T^2$ and the closed two-dimensional cross-cap.  One could
proceed as in section~\ref{sec:taub-cross}, quotienting
$[-\tstar,\tstar]\times T^3$ by the map $J:(t,x,y,z) \mapsto
(-t,x+\pi,y,z)$.  The only saddle points are then flat, and the
saddle-point action vanishes.  This could be interpreted as a classical
birth of a universe, if one were happy with the lack of time
orientability of this metric (and the concomitant implication that the
universe could die classically, in a time reversal of its birth).

\section{Summary and discussion}
\label{sec:discussion}

In this paper we have discussed the no-boundary path integral within
unimodular Einstein gravity in the Taub minisuperspace model with
$S^3$ spatial topology and in a Bianchi type I minisuperspace model
with $T^3$ spatial topology.  The (future-truncated) 4-manifolds
considered in the Taub model were the closed 4-dimensional ball and
the closed 4-dimensional cross-cap, while in the Bianchi type I model
we only considered the closed disk times~$T^2$.  In all three cases we
found a saddle point (or combination of them) for which the resulting
estimate to the wave function $\Psi$ is compatible with
normalizability and unitary evolution.  In the Bianchi type I model
the estimate was rapidly oscillating for $a^2b^2 \gg T$, and it
corresponded there to a family of isotropically expanding Lorentzian
universes. In the Taub model, on the other hand, the estimate did not
appear to have such a WKB region with either choice for the
4-manifold.

In the Taub model, with either 4-manifold, we also found a saddle
point for which the resulting estimate to the wave function is
compatible with normalizability and corresponds at late times to a
family of exponentially expanding Lorentzian universes.  However, both
these wave functions evolve nonunitarily, with probability (in the
sense of $|\Psi|^2$) being injected into the configuration space at an
exponentially increasing rate with respect to the unimodular time.

It should be emphasized that we did not attempt to define the path
integral beyond the saddle-point approximation.  In particular, we did
not attempt to discuss how good our saddle-point estimate of $\Psi$
should be expected to be.  
It would be possible to make some estimates on the pre-exponential
factor 
(assuming our choice of factor ordering in the Hamiltonian
operator), but this would seem to contain little information beyond
what we already have.  In particular, for the saddle points compatible
with unitary evolution, the saddle-point estimate does not seem to
contain enough information to single out a particular
selfadjoint extension of the Hamiltonian.

When specializing the saddle points of the Taub model to those of the
Friedmann model, we found, for each of the two 4-manifold topologies
we considered, one saddle point for which an unnormalizable Taub wave
function becomes a Friedmann wave function that is compatible with
normalizability, and even compatible with unitary evolution.  While
these saddle points would thus have seemed highly appealing in the
Friedmann model, the properties of interest disappear upon
generalization to the Taub anisotropy.  This should alert one to the
need to understand whether our results in the Taub model and the
Bianchi type I model would remain qualitatively unchanged upon the
addition of still more degrees of freedom.

One avenue towards investigating this would be to include some
inhomogeneous perturbations in the path integral as linearized
``test'' fields\cite{hall-hartle-con,jjhlou3}.  For example, if one
adds to the Friedmann model a massless scalar test field, and takes
for background the $\ballfourc$ saddle point metric with $\epsilon=1$
and $\eta=-1$, then one does not obtain a normalizable saddle point
wave function for the scalar field.  In this case, therefore, the
criterion of a normalizable scalar field perturbation wave function
around the Friedmann saddle point metric agrees with the criterion of
a normalizable Taub wave function.  

The underlying question here is how one can actually recognize which
saddle points, if any, yield a good approximation to the original path
integral (\ref{taub-nb-integral}).  In principle this reduces to the
question whether a given saddle point metric $g$ 
can be reached by deforming the
gravitational path integral from an originally Lorentzian ``contour''
to a complex contour passing through the saddle point in question.
For such a deformation to be valid, the path integral would, at a
minimum, have to be convergent for all intermediate values of the
contour, and one might, in a preliminary formulation, reduce this to
the question whether the complex metric $g$ can be reached by a curve
of metrics $g(s)$ all of which admit a convergent path integral for a
test scalar field (a type of perturbation that has much in common with
perturbations of the metric itself).  In Ref.\ \cite{yarmulke} 
a criterion of this type was used to fix the sign
of the imaginary part of a complex regulator that was there added to
the metric.  

Unfortunately, the issues raised in the previous two paragraphs are both
clouded by the fact that each of
our saddle points actually belongs to an entire family of saddle
points (all with the same action~$S$) whose members are related to
each other via ``complex diffeomorphisms'', or in other words
deformations of the complex path $t(s)$ that was used to parameterize
the general complex solution of the Einstein equations in
sections \ref{sec:taub-b4} and~\ref{sec:taub-cross}. 
Although $S$
itself does not depend on the choice of~$t(s)$, the criteria of
normalizable perturbation wave functions and convergent perturbation
path integrals apparently do.  We leave more systematic investigation
of these questions to the future.

The results we have just summarized cannot be claimed to be realistic,
of course, if only because they omit all other fields than gravity,
and because they represent situations of artificially imposed
symmetry.  Nevertheless, the saddle points we have found, and the
associated wave functions, contain enough interesting features to
warrant some further comments of an interpretive nature.

In order to be convincing, an interpretation of our results would have
to draw on a more general interpretive framework for quantum gravity
itself, in terms of which we could understand the significance of the
saddle point metrics and wave functions we have been computing.  
{}From 
a histories point of view, a quantum wave function has no direct
meaning at all.  Rather, it is seen as an intermediary in the
computation of the
{\em quantum measure\/}, 
$\mu(C)=||\psi_C||^2$, of the set $C$ of
Lorentzian manifolds (or more general histories) whose path integral
$\psi$ is.\footnote%
{In addition to its technical role as ``square root of the quantum
measure'', $\psi$ can serve as a summary of the past, useful for
computing the measures $\mu(C)$ of sets $C$ defined in the future.}
It is in terms of this quantum measure $\mu(C)$ 
[not to be confused with the ``measure-{\em factor\/}'' $d\nu(g)$ 
that occurs in expressions like $\int d\nu(g) e^{iS(g)}$] 
that predictions must be made.  
In special situations 
$\mu$ reduces to a probability, 
and more generally 
it seems to represent a kind of propensity 
for the actual history to belong to~$C$.   
In particular, 
one could probably interpret $|\psi(a,b;T)|^2$ 
in the present case 
as the probability density 
for the universe to find itself 
with the scale parameters $a$ and $b$ 
when the accumulated 4-volume reaches~$T$. 
(For more details see Refs.\
\cite{sorkin-talk,forks,sorkin-einstein,qmqmt,drexel}.)

Now, in nonrelativistic quantum mechanics, $\psi_C$ depends
parametrically on ordinary time $t$, and its squared norm
$||\psi_C||^2$ must be independent of $t$ in order that $\mu(C)$ be
defined consistently.  This independence is guaranteed by unitarity.
For gravity with $T=\,^4V$ playing the role of parameter time, an
analogous unitarity would seem to be required in order that the
quantum gravitational measure $\mu$ be well defined.  In our
minisuperspace truncation of general relativity, one can certainly
impose a unitary evolution on $\psi$ if one neglects topology change,
because the unimodular Hamiltonian operator is real in the
Schr{\"o}dinger representation, as pointed out earlier.  However, it
is not so clear how topology change affects the possibility of
unitarity, nor is it clear what is the proper class of spacetimes over
which the gravitational path integral should be taken in a
cosmological setting.  We believe that our results can shed some light
on both these questions.

One natural idea, suggested by what we know of the big bang, is that the
universe should expand from zero initial size.  In a discrete setting
this idea can perhaps be expressed by positing a single initial element
or ``origin'', in a continuous setting it must translate into conditions
on the topology and the metric of the spacetime manifold.  Let us assume
that the birth of a universe at zero size corresponds topologically to a
cobordism with empty initial boundary (it is thus a special case of
topology change).  References \cite{victoria,rafael-arvind,yarmulke}
delineate a class of 
symmetric 
tensor fields that exist on any cobordism,
that determine a well defined causal order, and that are globally smooth
with Minkowskian signature almost everywhere.  If we specialize them to
the case of a manifold appropriate to a big bang cosmology --- one
without initial boundary and compact toward the past --- then we arrive
at a 
{\em Lorentzian\/} 
version of the so called no-boundary proposal for
quantum cosmology
\cite{hawking-vatican,hartle-hawking,hawking-npb,hall-hartle-con}, with
a definite choice for the metrics to be integrated over.\footnote%
{Notice that the no-boundary condition, regarded in this manner, is a
 condition on the histories themselves; it need make no mention of any
 wave function. {}From a histories perspective, such a condition is
 more natural than any boundary condition couched in terms of
 the wave function~$\Psi$.}

In a cosmology of this sort, the path integral is a sum over certain
almost Lorentzian metrics on manifolds without initial boundary.  More
fundamentally, one might expect the sum to be over an underlying
discrete structure \cite{forks}, a possibility that could manifest
itself in the continuum in more than one way.  For example, it might
yield an amplitude for the universe to ``bounce'' (collapse and then
re-expand) or to develop into a ``bush-like'', multicomponent structure,
emerging from a single ``stem''.  In fact, some appeal to discreteness
might be required just for consistency: in order that the (approximate)
continuum wave function be truly square integrable (see below).

Now suppose for a moment that the only topology change that need be
considered is the initial expansion we have just been considering, and
that only a single macroscopic ``component'' of the universe comes
into being thereby.  
If we further fix the 4-manifold topology, then we are
left with a sum over almost Lorentzian 4-geometries on a given
manifold, and if the quantum measure $\mu(C)$ is to be formed in the
way suggested in \cite{sorkin-talk}, then for consistency, we need
$||\psi_C||^2$ to be independent of $T$ (for $T$ sufficiently large),
where in particular, $C$ can be the set of {\em all\/} 4-geometries
with $V=T$.  We also need, of course, that $||\psi||^2<\infty$ in
order that $\mu$ be defined at all.  To make contact with the analysis
in this paper, we need one further assumption, which is that the
Lorentzian functional integral defining $\psi$ can be analytically
continued to complex metrics and then approximated by deforming the
``integration contour'' to pass through a saddle point of the
analytically continued integrand.  This would justify (within a
minisuperspace truncation) the kind of approximate wave function we
have studied herein.  What is more, an analysis of the conditions of
validity for the contour deformation would tell us, in principle,
which saddle points (if any) actually contribute to
(\ref{taub-nb-estimate}), and with what signs.  
In particular, it would tell us whether our conditions of a smooth
complex metric on a smooth manifold without boundary correctly describe
the saddle points of the analytically continued functional integral.

The two key questions then are whether $\psi$ is $L^2$ (so that $\mu$
can be defined) and if so, whether $(\psi,\psi)$ is independent of $T$
(so that the definition can be consistent).  If the answer is yes,
then the picture painted above is at least internally coherent.
Lacking the deeper analysis that would tell us which saddle point(s)
we must use, the best we can say is that, for each of the 4-manifold
topologies we have studied in this paper, there is at least one saddle
point consistent with these two key features at the level of
approximation to which we are working.  In fact, there is precisely
one such saddle point for each manifold (actually a linear combination
of saddle points in the Taub cross-cap case), so we get in effect a
unique prediction of the no-boundary wave function in each case.

A further formal requirement, which would seem valid to the extent that
post-natal topology changes can be ignored, or at least localized at the
boundaries of configuration space, is that $\psi$ obey the unimodular
Schr{\"o}dinger equation~(\ref{schr-eq}).\footnote
{See, however, the doubts raised in 
Ref.\ \cite{claudio} 
about satisfaction of the Hamiltonian constraints in a 
path-integral 
formulation.}
We have seen that this demand is also met by all of our saddle points
(or appropriate combinations of them in the Taub-cross-cap case),
because, to the accuracy of our approximation, the Schr{\"o}dinger
equation reduces to the Hamilton-Jacobi equation, which all of our
saddle point actions satisfied.  Conversely,  the requirement would not
be met if we arbitrarily employed different saddle points for different
values of $T$, $a$, and~$b$.

Another formal issue on which we might have hoped for guidance from our
minisuperspace models is that of boundary conditions at the ``edge'' of
configuration space.  
If unitarity is to obtain 
then the only freedom in the boundary
conditions is that of a choice of selfadjoint extension for~$\hat{H}$.
But that ignores the possibility
of actual recollapse or, conversely, 
of a universe that remains ``pre-geometric'' 
for a long time $T$ and only then begins to expand.  
(For a causal set, the unimodular time $T$ would be identified with the
total number of elements \cite{forks,causet}.  
Even in a pre-geometric phase, therefore, $T$ retains its meaning.) 
A~selfadjoint
$\hat{H}$ also allows for recollapse, of course, but it demands then
an immediate ``bounce'', with no possibility of disappearance or of a
(temporary or permanent) transition to a disordered, non-geometric
phase.  Unfortunately, it is unlikely that our saddle point estimates
contain enough information to distinguish among these multiple
possibilities.  

Among our saddle points, there were ones exhibiting an exponential
decay of $|\psi|^2$ with time in every compact region of configuration
space.  This occurred, in particular, for every one of our ``unitary''
saddle points.  We chose to associate such decay with ``an escape
of probability toward~$\infty$'', but it might equally well signify an
escape through the finite boundary --- {\em i.e.}, a recollapse.
Although these two alternatives 
do not seem to be conclusively 
resolvable at our level of approximation,
the evidence points to a recollapse in the Taub cases and an unbounded
expansion in the Bianchi type I case.  
If this is correct, then the
evolution in the Taub cases is not actually unitary.

The wave functions $\Psi_-$ in the first Taub case and $\Psi_3$
in the second Taub case both look at $T=0$ qualitatively like
bound state wave functions localized near $uv=0$, except that they die
out exponentially as $T$ increases. 
The most direct interpretation of such behavior would
seem to be a universe expanding to Planckian size and then rapidly shrinking
to zero volume, at which point the saddle point approximation is helpless
to tell us what happens next.  
(Beyond the possibilities mentioned
earlier for what happens next, another is that probability escapes to
inhomogeneous metrics 
--- {\em i.e.}, 
the minisuperspace approximation breaks down.)\footnote
{If the universe really can die out entirely, then the rule 
 given in \cite{qmqmt} for
 forming the quantum measure $\mu(C)$ needs to be generalized in a way
 that allows paths to ``exit'' the configuration space at its
 boundary, without reappearing elsewhere.  This does appear to be
 possible, but only if one requires both halves of the ``Schwinger
 history'' to exit at the same place and at the same value of~$T$, which
 then serves as a premature ``truncation time'' for the exiting
 histories.}

In the Bianchi type I universe of section~\ref{sec:bianchiI}, 
the ``unitary'' wave function, called
there $\Psi_+$, behaves very differently.  For small $T$ (\ref{S-BI})
strongly resembles the action of a free nonrelativistic point particle
(albeit with an indefinite mass tensor) released from $u=v=0$ at
$T=0$, and we have noted in this connection that probability appears
to flow toward infinity.  This suggests that $\int{}dudv|\psi|^2$,
which here diverges marginally, would continue to be infinite in a
better approximation.  One might attribute this divergence either to a
failure of the no-boundary prescription, or to the idealization of
spacetime as a continuum, for which the saddle point approximation is,
disappointingly, not providing an effective cutoff.  If this is
correct, then to compensate, we might need to invoke discreteness
explicitly, by smearing $\psi$ out by hand from a delta-function to
Planckian size at $T=0$.  This in turn
would be expected to modify its
behavior for $uv\gg{T^2}$.

We also found saddle points (and with many attractive features) that
manifested an exponential {\em growth\/} of norm. For them,
probability seems to be flowing {\em inward\/} from the finite
boundary at an increasing rate.  The nearest we can come to a picture
of what this type of unitarity breakdown might mean would be a
pre-geometric universe continually sending out branches that develop
into continuum spacetimes.  Such a ``bush-like'' universe could not
really correspond to a one-configuration-space wave function at all,
though, and it is not really clear whether any plausible
interpretation can accommodate such saddle points.

A question often raised in connection with a quantum cosmological
model is whether it predicts that the universe will, at late times,
expand along an approximately classical trajectory.  
In effect, one is
asking whether the universe, 
having once arrived at certain values of $T$, 
$a$, and~$b$, 
can be expected to continue its expansion along some
particular classical trajectory through these values.  In the
affirmative case, we may say that it makes a spontaneous transition to
classical behavior, possibly after some initial era of non-classical
expansion.\footnote
{By saying that a history (Lorentzian geometry) evolves
 ``non-classically'' we merely mean that it fails to satisfy the
 classical Einstein equations.}
One can argue that such behavior is correlated with rapid, WKB-like
oscillations of $\psi$ in the region in question.  Another familiar
criterion for classical evolution is the validity of a stationary
phase approximation to the path integral, meaning that most of the
contribution to the quantum measure comes from nearly classical
histories, and this in turn can be related to the saddle point
metric's being 
(up to ``complex diffeomorphism'')
nearly Lorentzian at late times.  To us, it seems
unclear whether any of these features is really necessary or whether
decoherence effects associated with gravitons or other matter could by
themselves bring about a transition to a nearly classical universe
like the one we inhabit.  At any rate, it would seem helpful at least
for the universe to attain a large size with high probability, and
this happens (together with a spontaneous transition to classical
expansion if $T\ll{}a^2b^2$) in at least one of our normalizable and
unitary cases, namely for the $\eta=+1$ saddle point in our Bianchi 
type I model.

As pointed out earlier, it is meaningful in the unimodular theory to
ask what the wave function looks like at $T=0$, the ``moment of
birth'', and consistency would seem to require that the no-boundary
prescription yield a universe which is of zero or Planckian size.
This was the case for all of our ``unitary'' saddle points, so to that
extent, the desired consistency seems to be present.  (In contrast, the
non-unitary but normalizable 
wave function $\Psi_0$ in the Taub
cross-cap case seems to look at $T=0$ like a combination of a
delta-function with a ``zero momentum'' state spread out over all of
configuration space.  We recall that this $\Psi$ resulted
for $\alpha<\alpha_c$ from a time-non-orientable, purely Lorentzian
solution, which one suspects is not a valid saddle point at all.)

In concluding, it seems fitting to remark on the rather lifelike nature
of some of our models.  In contrast to non-unimodular versions of
quantum cosmology, where the wave function is typically non-normalizable
and otherwise very hard to interpret, we have found here many saddle
points yielding $\psi$'s which are either $L^2$ or marginally so, and
which evolve consistently with unitarity at our degree of approximation.
This seems encouraging for the account of topology change sketched
above.

Among all of our saddle points, there is precisely one that is
consistent with normalizability and unitarity and that spontaneously
makes a transition to classical expansion.  Interestingly, it belongs to
spatial topology $T^3$ and not to~$S^3$.  In this way, we might imagine
predicting something about the large scale topology of the universe, if
it turned out that this distinction between $T^3$ and $S^3$ persisted in
more realistic models.

\acknowledgments
We are grateful to
Abhay Ashtekar,
Alan Coley,
John Friedman,
Andrei Linde,
Don Marolf, 
Michael Ryan,
and especially Gu Zhichong
for discussions.
This work was supported in part by NSF grants
PHY90-05790,
PHY90-16733,
PHY91-05935,
PHY-94-21849,
and 
PHY-96-00620,   
and by research funds provided by Syracuse University.
The work of one of us (J.L.) was done in part 
during a leave of absence from 
Department of Physics, 
University of Helsinki.

\newpage



\begin{references}


\bibitem{sorkin-talk}
R.~D. Sorkin,
 ``On the Role of Time in the Sum-over-histories Framework for Gravity'',
 paper presented to the conference on
 The History of Modern Gauge Theories,
 held Logan, Utah, July 1987,
 published in
Int.\ J. Theor.\ Phys.\ {\bf 33}, 523 (1994).

\bibitem{goa}
R.~D. Sorkin, 
 ``A Modified Sum-Over-Histories for Gravity'',
   reported in
  {\it 
   Proceedings of the International Conference on Gravitation and Cosmology, 
   Goa, India, 14-19 December, 1987\/},
   edited by 
   B.~R. Iyer, A.~Kembhavi, J.~V. Narlikar, and C.~V. Vishveshwara,
   see pages 184-186 in the article by 
   D.~Brill and L.~Smolin: 
   ``Workshop on quantum gravity and new directions'', pp.\ 183-191 
   (Cambridge University Press, Cambridge, England, 1988). 

\bibitem{forks}
R.~D. Sorkin,
Int.\ J.\ Theor.\ Phys.\ {\bf 36}, 2759 (1997). 
(gr-qc/9706002)

\bibitem{einstein}
A.~Einstein,
Sitzungsber.\ K. Preuss.\ Akad.\ Wiss.
349 (1919);
English translation in {\it The Principle of Relativity\/}
(Methuen, 1923;
reprinted by Dover, New York, 1952), p.~189. 

\bibitem{vanderBij}
J.~J. van der Bij, H.~van Dam, and Y.~J. Ng,
Physica (Utrecht) {\bf 116A}, 307 (1982).

\bibitem{wilczek}
F.~Wilczek,
Phys.\ Rep.\ {\bf 104}, 111 (1984).

\bibitem{zee}
A.~Zee, in
{\it Gauge Interactions, Proceedings of the 
Twentieth Orbis Scientiae},
Miami, Florida, 1983, edited by 
B.~Kursunoglu, 
S.~L. Mintz, and 
A.~Perlmutter
(Plenum, New York, 1984).

\bibitem{buchdragon}
W.~Buchm\"uller and N.~Dragon,
Phys.\ Lett.\ B {\bf 207}, 292 (1988).

\bibitem{weinberg}
S.~Weinberg,
Rev.\ Mod.\ Phys.\ {\bf 61}, 1 (1989).

\bibitem{unruh}
W.~G. Unruh,
Phys.\ Rev.\ D {\bf 40}, 1048 (1989);
in {\it Quantum Gravity}, 
Proceedings of the Fourth Seminar on Quantum Gravity, 
Moscow, USSR, 1987,
edited by 
M.~A. Markov, 
V.~A. Berezin, and 
V.~P. Frolov 
(World Scientific, Singapore, 1988).

\bibitem{unruh-wald}
W.~G. Unruh and
R.~M. Wald,
Phys.\ Rev.\ D {\bf 40}, 2598 (1989).

\bibitem{henn-teitel}
M.~Henneaux and C.~Teitelboim,
Phys.\ Lett.\ B {\bf 222}, 195 (1989).

\bibitem{brown-york}
J.~D. Brown and
J.~W. York,
Phys.\ Rev.\ D {\bf 40}, 3312 (1989).

\bibitem{bombelli-banff}
L.~Bombelli,
in {\it Gravitation --- A Banff Summer Institute}, 
edited by R.~Mann and P.~Wesson
(World Scientific, Singapore, 1991).

\bibitem{kuchar}
K.~V. Kucha\v{r},
Phys.\ Rev.\ D {\bf 43}, 3332 (1991).

\bibitem{casta-lom}
M.~A. Castagnino and
F.~Lombardo,
Phys.\ Rev.\ D {\bf 48}, 1722 (1993).

\bibitem{feyn}
R.~P. Feynman,
Rev.\ Mod.\ Phys.\ {\bf 20}, 367 (1948). 

\bibitem{griffiths}
R.~Griffiths,
J.~Stat.\ Phys.\ {\bf 36}, 219 (1984). 

\bibitem{sorkin-einstein}
R.~D. Sorkin,
in
{\it Conceptual Problems of
Quantum Gravity\/}
(Einstein Studies~II), 
edited by A.~Ashtekar and J.~Stachel
(Birkhauser, Boston, 1991).

\bibitem{ishamhist}
C.~J. Isham, 
J.~Math.\ Phys.\ {\bf 35}, 2157 (1994). 
(gr-qc/9308006)

\bibitem{qmqmt}
R.~D. Sorkin,
Mod.\ Phys.\ Lett.\ A {\bf 9}, 3119 (1994). 
(gr-qc/9401003)

\bibitem{hartlehist}
J.~B. Hartle, 
 in 
 {\it Gravitation and Quantizations, 
Les Houches, session LVII, 1992\/},
edited by B.~Julia and J.~Zinn-Justin 
(Elsevier Science B.V., Amsterdam, 1995). 
(gr-qc/9304006)

\bibitem{drexel}
R.~D. Sorkin,
in 
{\it Proceedings of the Fourth Drexel Symposium on Quantum
       Nonintegrability: Quantum Classical Correspondence},
       held Philadelphia, September 8-11, 1994,  
       edited by D.~H. Feng and B-L. Hu 
    (International Press, Cambridge, Massachusetts, 1997), 
    pp.\ 229--251. 
    (gr-qc/9507057)

\bibitem{ash-time}
{\it Conceptual Problems of
Quantum Gravity\/}
(Einstein Studies~II), 
edited by A.~Ashtekar and J.~Stachel
(Birkhauser, Boston, 1991).

\bibitem{kuchar-winn}
K.~V. Kucha\v{r}, in
{\it Proceedings of the 4th 
Canadian Conference on General Relativity
and Relativistic Astrophysics\/},
edited by 
G.~Kunstatter, 
D.~E. Vincent and 
J.~G. Williams
(World Scientific, Singapore, 1992).

\bibitem{isham-time}
C.~J. Isham,
in {\it Integrable Systems, Quantum Groups and
Quantum Field Theory\/},
edited by L.~A. Ibart and M.~A. Rodrigues
(Kluwer, Dordrecht, 1992).
(gr-qc/9210011)

\bibitem{hawking-vatican}
S.~W. Hawking, 
in {\it Astrophysical Cosmology: 
Proceedings of the Study
Week on Cosmology and Fundamental Physics,} 
edited by 
H.~A. Br\"uck, 
G.~V. Coyne and 
M.~S. Longair
(Pontificiae Academiae Scientiarum Scripta Varia,
Vatican City, 1982).

\bibitem{hartle-hawking}
J.~B. Hartle and S.~W. Hawking,
Phys.\ Rev.\ D {\bf 28}, 2960 (1983).

\bibitem{hawking-npb}
S.~W. Hawking,
Nucl.\ Phys.\ {\bf B239}, 257 (1984).

\bibitem{linde-zetf}
A.~Linde,
Zh.\ Eksp.\ Teor.\ Fiz.\ {\bf 87}, 369 (1984)
[Sov.\ Phys.\ JETP {\bf 60}, 211 (1984)].

\bibitem{linde-ncim}
A.~Linde,
Lett.\ Nuovo Cimento {\bf 39}, 401 (1984).

\bibitem{linde-repprog}
A.~Linde,
Rep.\ Prog.\ Phys.\ {\bf 47}, 925 (1984).

\bibitem{linde-mezh}
A.~Linde and A.~Mezhlumian,
Phys.\ Lett.\ B {\bf 307}, 25 (1993).
(gr-qc/9304015)

\bibitem{linde2-mezh}
A.~Linde, D.~Linde, and A.~Mezhlumian,
Phys.\ Rev.\ D {\bf 49}, 1783 (1994).
(gr-qc/9306035)

\bibitem{vile1}
A.~Vilenkin,
Phys.\ Rev.\ D {\bf 30}, 509 (1984).

\bibitem{vile2a}
A.~Vilenkin,
Phys.\ Rev.\ D {\bf 33}, 3560 (1986).

\bibitem{vile2b}
A.~Vilenkin,
Phys.\ Rev.\ D {\bf 37}, 888 (1988).

\bibitem{vile3}
A.~Vilenkin,
Phys.\ Rev.\ D {\bf 50}, 2581 (1994).
(gr-qc/9403010) 

\bibitem{vile-erice}
A.~Vilenkin,
in {\it String Gravity and Physics at the Planck Energy Scale\/}
(NATO Advanced Study Institute,
 series~C: Mathematical and physical sciences, Vol.~476),
edited by N.~Sanchez and A.~Zichichi
(Kluwer Academic, Dordrecht, Netherlands, 1996).
(gr-qc/9507018)

\bibitem{vile-discord}
A.~Vilenkin, 
``The wave function discord'', 
gr-qc/9804051. 

\bibitem{victoria}
R.~D. Sorkin, 
  in 
  {\it Proceedings of the Third Canadian Conference on 
     General Relativity and Relativistic Astrophysics} 
  (held Victoria, Canada, May 1989), 
  edited by A.~Coley, F.~Cooperstock, and B.~Tupper
  (World Scientific, Singapore 1990),
  pp.\ 137-163. 

\bibitem{rafael-arvind}
A.~Borde and R.~D. Sorkin,  
``Causal Cobordism: Topology Change Without Causal Anomalies''
(in preparation). 

\bibitem{hawkingCC}
S.~W. Hawking,
in {\it General Relativity: An Einstein Centenary Survey,}
edited by S.~W. Hawking and W.~Israel
(Cambridge University Press, Cambridge, 
England, 1979).

\bibitem{hall-hartle-con}
J.~J. Halliwell and
J.~B. Hartle,
Phys.\ Rev.\ D {\bf 41}, 1815 (1990).

\bibitem{jjhlou3}
J.~J. Halliwell and
J.~Louko,
Phys.\ Rev.\ D {\bf 42}, 3997 (1990).

\bibitem{ryan-shepley}
M.~P. Ryan and 
L.~C. Shepley, 
{\it Homogeneous relativistic cosmologies\/}
(Princeton Univerity Press, Princeton, New Jersey, 1975).

\bibitem{jantzen-cmp}
R.~T. Jantzen, 
Commun.\ Math.\ Phys.\ {\bf 64}, 211 (1979).

\bibitem{exact}
D.~Kramer, H.~Stephani, E.~Herlt, and
M.~MacCallum, {\it Exact Solutions of Einstein's Field Equations,}
edited by E.~Schmutzer (Cambridge University Press, Cambridge,
England, 1980), Section 11.3.1.

\bibitem{rayrev}
R.~Laflamme,
in
{\it Quantum Gravity and Cosmology: Proceedings of the XXII GIFT
International Seminar on Theoretical Physics,}
edited by J.~P\'erez-Mercader,
J.~Sol\`a and E.~Verdaguer
(World Scientific, Singapore, 1992).

\bibitem{mike-rds} 
M.~P. Ryan and R.~D. Sorkin, 
``On Factor Ordering and the use of Spacetime Volume 
as a Time Parameter''
(in preparation)

\bibitem{reed-simonII}
M.~Reed and B.~Simon,
{\it Methods of Modern Mathematical Physics\/}
(Academic, New York, 1975),
Vol.~II.

\bibitem{yarmulke}
J.~Louko and  R.~D. Sorkin,
Class.\ Quantum Grav.\ {\bf 14}, 179 (1997). 
(gr-qc/9511023)

\bibitem{jlr}
L.~G. Jensen, J.~Louko, and P.~J. Ruback,
Nucl.\ Phys.\ {\bf B351}, 662 (1991).

\bibitem{non-t-orient}
R.~D. Sorkin, 
Int.\ J. Theor.\ Phys.\ {\bf 25}, 877 (1986). 

\bibitem{fried-higu}
J.~L. Friedman and A.~Higuchi,
Phys.\ Rev.\ D {\bf 52}, 5687 (1995).
(gr-qc/9505035)

\bibitem{dls-waterloo}
A.~Daughton, J.~Louko, and
R.~D. Sorkin,
in {\it Proceedings of the 5th Canadian Conference on
General Relativity and Relativistic Astrophysics\/},
edited by R.~B. Mann and R.~G. McLenaghan
(World Scientific, Singapore, 1994).
(gr-qc/9305016)

\bibitem{wolf}
J.~A. Wolf,
{\it Spaces of Constant Curvature\/}
(McGraw-Hill, New York, 1967).

\bibitem{ellis}
G.~F.~R. Ellis,
Gen.\ Relativ.\ Gravit.\ {\bf 2}, 7 (1971). 

\bibitem{louko-ruback}
J. Louko and P.~J. Ruback,
Class.\ Quantum Grav.\ {\bf 8}, 91 (1991).

\bibitem{giu-lou-thetas}
D.~Giulini and
J.~Louko,
Phys.\ Rev.\ D {\bf 46}, 4355 (1992).
(hep-th/9203007)

\bibitem{hoso+1}
T.~Koike, M.~Tanimoto, and A.~Hosoya,
J.\ Math.\ Phys.\ {\bf 35}, 4855 (1994).
(gr-qc/9405052)

\bibitem{waelbroeck}
H.~Waelbroeck, 
Commun.\ Math.\ Phys.\ {\bf 170}, 63 (1995). 
(gr-qc/9311033)

\bibitem{hoso+2}
M.~Tanimoto, T.~Koike, and A.~Hosoya,
J.\ Math.\ Phys.\ {\bf 38}, 350 (1997).
(gr-qc/9604056)

\bibitem{hoso+3}
M.~Tanimoto, T.~Koike, and A.~Hosoya,
J.\ Math.\ Phys.\ {\bf 38}, 6560 (1997).
(gr-qc/9705052) 

\bibitem{kodama-homog}
H.~Kodama,
Prog.\ Theor.\ Phys.\ {\bf 99}, 173 (1998). 
(gr-qc/9705066)

\bibitem{laf-stwo}
R.~Laflamme,
in 
{\it Origin and Early History of the Universe:
Proceedings of the 26th Li\`ege International
Astrophysical Colloquium (1986)\/},
edited by
J.~Demaret
(Universit\'e de Li\`ege,
Institut d'Astro\-physique, 1987);
Ph.D. thesis (University of Cambridge,
1988).

\bibitem{lou-annphys}
J.~Louko,
Ann.\ Phys.\ (N.Y.) {\bf 181}, 318 (1988);
Class.\ Quantum Grav.\ {\bf 8}, 1947 (1991).

\bibitem{lou-canoniz}
J.~Louko,
Phys.\ Lett.\ B {\bf 202}, 201 (1988).

\bibitem{claudio}
 C.~Teitelboim, 
Phys.\ Rev.\ Lett.\ {\bf 50}, 705 (1983). 

\bibitem{causet}
L.~Bombelli, J.~Lee, D.~Meyer, and R.~D. Sorkin, 
Phys.\ Rev.\ Lett.\ {\bf 59}, 521 (1987). 


\end{references}
\end{document}